\documentclass{mn2e}

\usepackage{amsfonts}
\usepackage{amsmath}
\usepackage{graphicx}

\title[Rank-ordered marks]
      {Measures of galaxy environment -- II. Rank-ordered mark correlations}
\author[R. Skibba et al.]
 {Ramin A. Skibba$^{1,2}$\thanks{E-mail: rskibba@ucsd.edu}, 
Ravi K. Sheth$^{3,4}$, Darren J. Croton$^5$, Stuart I. Muldrew$^6$, \newauthor 
Ummi Abbas$^7$, Frazer R. Pearce$^6$, Genevieve M. Shattow$^5$\\
  $^{1}$Steward Observatory, University of Arizona, 933 N. Cherry Ave., Tucson, AZ 85721, USA\\
  $^{2}$Center for Astrophysics and Space Sciences, Department of Physics, University of California, 9500 Gilman Dr., La Jolla, San Diego, CA 92093, USA\\
  $^{3}$The Abdus Salam International Center for Theoretical Physics, Strada Costiera 11, 34151 Trieste, Italy\\
  $^{4}$Center for Particle Cosmology, University of Pennsylvania, 209 S. 33rd St., Philadelphia, PA 19104, USA\\
  $^{5}$Centre for Astrophysics \& Supercomputing, Swinburne University of Technology, P.O. Box 218, Hawthorn, VIC 3122, Australia\\
  $^{6}$School of Physics and Astronomy, University of Nottingham, Nottingham, NG7 2RD, UK\\
  $^{7}$INAF - Osservatorio Astrofisico di Torino, 10025 Pino Torinese, Italy
}

\newcounter{appfig}

\begin{document}

\pagerange{\pageref{firstpage}--\pageref{lastpage}}

\maketitle
\label{firstpage}

\begin{abstract}
%here are our main conclusions... 
We analyze environmental correlations using mark clustering statistics with the mock galaxy catalogue constructed by Muldrew et al.\ (Paper I).  
We find that mark correlation functions are 
%particularly sensitive to environmental trends and are 
able to detect even a small dependence of galaxy properties on the environment, quantified by the overdensity $1+\delta$, while such a small dependence would be difficult to detect by traditional methods.  
We then show that rank ordering the marks and using the rank as a weight is a simple way of comparing the correlation signals for different marks.  
With this we quantify to what extent fixed-aperture overdensities are sensitive to large-scale halo environments, nearest-neighbor overdensities are sensitive to small-scale environments within haloes, and colour is a better tracer of overdensity than is luminosity.
\end{abstract}


\section{Introduction}\label{sec:intro}

In hierarchical clustering models, clues about the galaxy formation 
process are encoded in correlations between galaxy properties and 
their environments.  This has motivated measurements of such 
correlations.  Traditional measures are intended to allow one to 
quantify if galaxies in dense regions tend to be more luminous, or 
redder, or older, or tend to move faster than average, and so on.  
These conclusions depend critically on how the density $N_g/V$ was 
estimated: fixed aperture measurements count the number of galaxies 
$N_g$ that are within volume $V$ of an object (i.e., the numerator 
of the ratio $N_g/V$ varies from one object to another), 
whereas near-neighbour measurements find the $V$ that contains 
$N_g$ nearest neighbours (i.e., the denominator is stochastic).  
Clearly, the size and shape of $V$, or the choice of $N_g$ matter 
greatly (the universe is homogeneous on sufficiently large $N_g$ or $V$).  
%Frazer: In the introduction, the distinction between true 3-d measures (or 2+1d) and 2d measures wasn't imediately clear. For instance, what does "volume" mean here? Is Ng a 2d, a 2+1d or a 3d measure? Is it "density" or "surface density" etc. I realise the answer is "depends which measure you take" and that it is hard to be concise.
In addition, the choice of three-dimensional or projected surface density matters as well, 
as do the redshift uncertainties and sample selection. 
Determining which of the many observed correlations is 
fundamental, and which is a consequence of others, can be a subtle 
task, especially since the environment is often the least well determined 
of a galaxy's attributes.  Moreover, the estimate of the environment 
is often sufficiently complicated that it cannot be modelled or 
interpreted analytically.  

Mark clustering statistics are fundamentally different, in the sense that they are, 
strictly speaking, statements about pairs, triples, quadruples, 
etc., of galaxies, rather than about single objects 
(Stoyan \& Stoyan 1994).  For example, 
the most commonly used such statistic returns an estimate of how the 
properties of galaxy pairs (rather than of single galaxies) depend on 
pair separation.  (While this is easily extended to triples, quadruples, 
etc., such estimates are rarely ever made.)  
In essence, for each pair separation $r$, this statistic weights 
each galaxy in a pair by its own attribute (e.g., luminosity, colour, 
etc., expressed in units of the mean across the population) and then 
divides this weighted pair count by the unweighted one.  
Symbolically, one may write this statistic as $WW(r)/DD(r)$, where 
$WW$ and $DD$ stand for the weighted and unweighted pair counts at 
separation $r$.  

Previous estimates of $WW/DD$ have shown that close pairs of galaxies 
are more luminous (Beisbart \& Kerscher 2000), redder (Skibba et al. 2006), 
and older (Sheth et al. 2006) than average ($WW/DD > 1$ for $r$ less 
than a few $h^{-1}$Mpc).  
While these trends are qualitatively the same as those returned by 
traditional estimates, and they are also in qualitative agreement 
with galaxy formation models (Sheth 2005), the mark statistics are 
particularly interesting because a theoretical framework exists for 
interpreting such measurements quantitatively (Sheth 2005; 
Skibba et al. 2006).  
On the other hand, this is also a drawback, because the theoretical 
framework is almost required if one wishes to draw more than qualitative 
conclusions from such measurements.  This is because the magnitude of 
the (say) luminosity-weighted signal changes if one weights instead by 
the log of the luminosity (Sheth, Connolly \& Skibba 2005; 
Skibba et al.\ 2006).  Since the same physics has led to both signals, 
one would like the measurement to not depend on the `units' in which 
the measurement was made.  

This dependence on `units' derives from the fact that the magnitude 
of $WW/DD$ depends on the distribution of the weights (e.g., its 
width, the length of its tails, etc.).  Needless to say, it also 
complicates efforts to determine if one observable correlates more 
strongly with environment than another.  As a case in point, it has 
long been known that cluster galaxies tend to be redder than average, 
but there is a wide range in luminosity between the brightest cluster galaxy (BCG) 
and the dwarf satellites in a cluster.  Since clusters are regions of high density, 
one naively expects to find that colour correlates more strongly with 
environment than does luminosity.  
However, the mark correlation signal appears to have a larger 
amplitude for luminosity than it does for colour (Skibba \& Sheth 2009).  
One of the main goals of the present work is to show how to remove 
this effect from the measurement, so that the magnitude of the signal 
can be compared across different weights.  

In the next section, we describe the galaxy catalogues used throughout the paper. 
In Section~\ref{sec:MCFresults} we demonstrate that mark correlations are particularly 
sensitive probes of environmental correlations; they correctly show a 
strong signal even when traditional estimates based on how one-point 
statistics vary as a function of environment are unable to see one -- a 
fact that was recently exploited by Paranjape \& Sheth (2012).  
In Section~\ref{sec:markstats} we show how to remove the effect of the mark correlation 
signal on `units', arguing that for any given weight, one should simply 
rank order and use the rank as the weight.  %rescale to the interval [0,1].  
We then use these rank-ordered mark correlations to compare different 
mark correlation signals with one another.  
A final section summarizes our findings.

%Throughout, we show measurements in the main galaxy sample of SDSS DR7, 
%volume limited to $M_r < -19$.  To illustrate our methods, and to 
%interpret some of our results, we make similar measurements in a 
%publicly available mock catalog (Muldrew et al. 2012, hereafter M12), 
%which used the halo-model decomposition of the luminosity and color 
%dependence of clustering in the SDSS (following the algorithm 
%described in Skibba \& Sheth 2009).  

Throughout this paper we assume a spatially flat cosmology with 
$\Omega_m=0.25$ and $\Omega_\Lambda=0.75$, and $\sigma_8=0.9$. 
We write the Hubble constant as $H_0=100h$~km~s$^{-1}$~Mpc$^{-1}$.

%Introduce: density/environment dependence of haloes and galaxies in 
%the current paradigm of $\Lambda$CDM hierarchical structure formation 
%and galaxy formation.

%Describe models connecting galaxies to haloes: HOD/CLF/HAM and 
%observationally constrained SAMs/simulations. 

%Briefly describe different measures of galaxy environment.

%In Muldrew et al.\ (2012; hereafter M12), we analyzed a variety of 
%galaxy environment measures in the literature, using a singly mock 
%galaxy catalogue, based on a halo-model description of the 
%distribution and clustering of galaxy luminosities and colours in 
%the SDSS.

%This paper is organized as follows. 
%In the next section, we briefly introduce mark clustering statistics, 
%and focus on the mark two-point correlation function. %\ref{sec:markstats}
%In Section~\ref{sec:mock}, we describe the mock galaxy catalogue used in this analysis, 
%and we describe the environment measures in Section~\ref{sec:enviromeas}. 
%We present our results in Sections~\ref{sec:MCFresults} and \ref{sec:dists}, 
%showing the environmental sensitivity of mark correlation functions. 
%Finally, we summarize the results and discuss their implications in 
%Section~\ref{sec:conc}. %and \ref{sec:halomodel}

%[should this section be shortened?]

% p(<V|N) = p(>N|V)
% p(V) = d/dV p(>N|V)
% rho = N/V
% p(rho)drho = p(V)dV
% rho p(rho) = Vp(V) (dV/V rho/drho)
% for poisson point clusters:
% p(N|V) = 

\section{Data}

\subsection{Mock Galaxy Catalogue}\label{sec:mock}
%To illustrate our methods, and to 
%interpret some of our results, we make similar measurements in a 
%publicly available mock catalog (Muldrew et al. 2012, hereafter M12), 
%which used the halo-model decomposition of the luminosity and color 
%dependence of clustering in the SDSS (following the algorithm 
%described in Skibba \& Sheth 2009).  

To illustrate our methods and to interpret some of our results, we use the 
mock galaxy catalogue of Muldrew et al. (2012; hereafter M12).  We refer the 
reader to M12 for details about the dark matter simulation, halo-finding 
algorithm, and the procedure for populating the haloes with galaxies.

We begin with the Millennium Simulation (Springel et al.\ 2005), which is a large $N$-body simulation of dark matter structure in a cosmological volume.  
Dark matter particles are traced in a cubic box of $500 h^{-1} {\rm Mpc}$ on a side, with a halo mass resolution of $\sim5\times10^{10}h^{-1}M_\odot$.  Collapsed haloes with at least 20 particles are identified with a friends-of-friends group finder.

The haloes are populated with galaxies with luminosities and colours, following the 
algorithm described in Skibba et al.\ (2006) and Skibba \& Sheth (2009), which is 
constrained by the luminosity and colour distribution and clustering in the Sloan Digital Sky Survey (SDSS; York et al.\ 2000). 
An important assumption in the model is that all galaxy properties---their numbers, spatial distributions, velocities, luminosities, and colours---are determined by halo mass alone.  
We specify a minimum $r$-band luminosity for the galaxies in the catalogue, $M_r=-19$, to stay well above the resolution limit of the Millennium Simulation, avoiding any issues of completeness that may bias our results.  

\begin{figure}
 \includegraphics[width=\hsize]{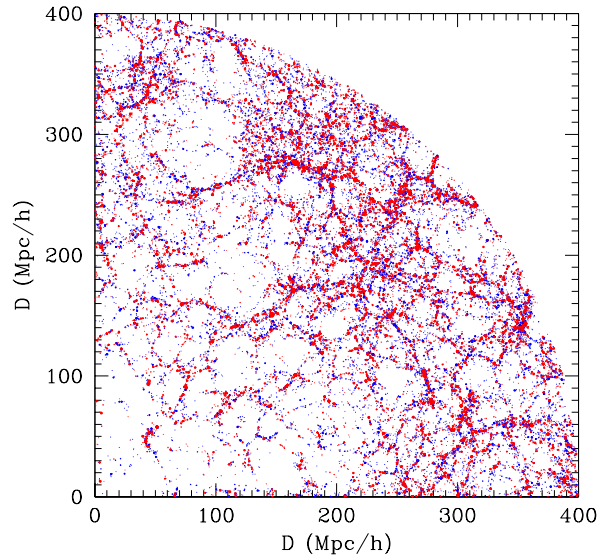}%{mockconesliceplot_4deg.ps}%mockconesliceplot_blackbg_4deg.ps
 \caption{Slice of the redshift-space mock light cone, showing galaxies within $\pm4$ deg. %of the celestial equator.
          Red/blue points are galaxies on the red/blue sequences of the colour-magnitude diagram, 
          and larger/smaller points are brighter/fainter galaxies.}
 \label{fig:slice}
\end{figure}

This procedure produces a mock galaxy catalogue containing 1.84 million galaxies, of which 29 percent are `satellite' galaxies.  Galaxies occupy haloes with masses ranging from $10^{11}$ to $10^{15.3}h^{-1} M_\odot$.  
We show a slice of the mock light cone in Figure~\ref{fig:slice}, in which more luminous galaxies are identified with larger points, and red and blue sequence galaxies with red and blue points. 

\subsection{SDSS Galaxy Catalogue}\label{sec:sdss}

For comparison, we will also show clustering measurements in the main galaxy sample of 
SDSS Seventh Data Release (DR7; Abazajian et al.\ 2009), in a catalogue volume-limited to $M_r < -19$, with $0.02<z<0.0642$. 

\begin{figure}
 \includegraphics[width=\hsize]{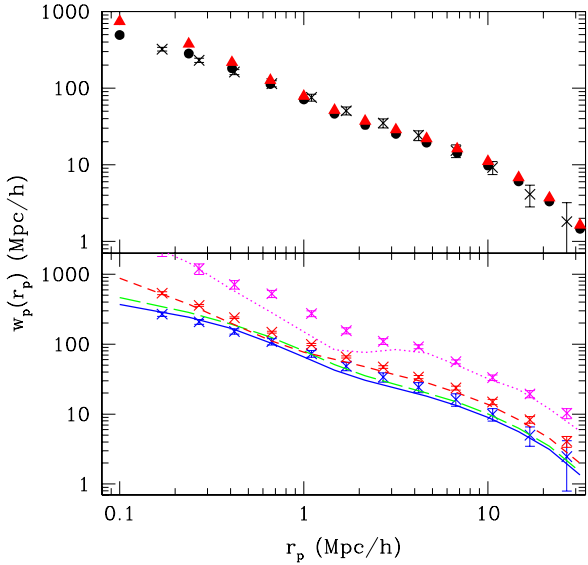}%{mockvssdss_Mr19_LdepCFs.ps} %{mock_Mr19_LdepCFs.ps}
 \caption{Upper panel: projected correlation function (circle points) and $r$-band luminosity-weighted correlation function (triangles) of the full $M_r<-19$ mock catalogue. 
          Lower panel: CFs for luminosity bins $-20<M_r<-19$ (blue solid line), $-21<M_r<-20$ (green long-dashed line), $-22<M_r<-21$ (red short-dashed line), and $-23<M_r<-22$ (magenta dotted line). 
          The crosses are the CFs measured from SDSS DR7 (Zehavi et al.\ 2011), and the measurement for $-21<M_r<-20$ is omitted, for clarity. }
 \label{fig:2PCF}
\end{figure}

Clustering measurements of galaxy redshift surveys have traditionally been done by splitting 
a catalogue in luminosity bins (e.g., Norberg et al.\ 2002; Zehavi et al.\ 2005, 2011; Li et al.\ 2006; Coil et al.\ 2008). 
We compare such clustering\footnote{The clustering measurements in this paper are performed using the $N$tropy code developed by Gardner, Connolly, \& McBride (2007).} 
in the SDSS and in the mock catalogue in Figure~\ref{fig:2PCF}. 
There is generally very good agreement, except at $0.7<r_p<3\,\mathrm{Mpc}/h$ for the brightest galaxies.  
Note that the mock catalogue was constrained by earlier SDSS data sets (Zehavi et al.\ 2005; Skibba et al.\ 2006), while these new measurements use the full DR7 data set (Zehavi et al.\ 2011). 

Most of the analysis of environmental correlations throughout this paper is based on the mock galaxy catalogue, with additional comparisons to the SDSS galaxy catalogue in Section~\ref{sec:LCrankMCFs}.

\section{Sensitivity to environmental correlations}\label{sec:MCFresults}

%\textbf{Do we need more intro to mark corrs?} 
In what follows, we will refer to any property of a galaxy, e.g., its luminosity or its colour, as a `mark'.  
As stated above, the most commonly used mark statistic is the mark correlation function, which is defined as the ratio of the weighted/unweighted correlation function:
\begin{equation}
  M(r) \,\equiv\, \frac{1+W(r)}{1+\xi(r)}\approx \frac{WW(r)}{DD(r)},
 \label{markedXi}
\end{equation}
where $WW/DD$ is the pair count ratio. 
The mark projected correlation function is similarly defined: 
$M(r_p)\equiv(1+W_p/r_p)/(1+w_p/rp)$. 
If the weighted and unweighted clustering are significantly different at a particular separation $r$, then the mark is correlated (or anti-correlated) with the environment at that scale;
the degree to which they are different quantifies the strength of the correlation.

In this section, we illustrate that mark correlation functions are particularly sensitive to environmental effects.  We do so by introducing a small additional dependence of a galaxy mark $w$ (luminosity or colour here) on the galaxy's overdensity (using one of the environment measures defined below).  (We use the letter $w$ because, when we discuss mark correlations below, we treat the mark $w$ as a `weight'.)  That is, if a galaxy has mark $w$, then we change it to
\begin{equation}
 w_\alpha \,=\, w\, (1+\delta)^\alpha .
 \label{eq:weight}
\end{equation}
We then rank order the marks $w_\alpha$ and rescale them so that they have the same distribution as before the environmental effect was added.  I.e., we require
\begin{equation}
 p(>w_\alpha) = p(>w).  
 \label{eq:rescale}
\end{equation}
Therefore, by construction, there is no trace of the additional correlation with environment in the one-point statistics of $w_\alpha$; it is only by studying how the one-point distribution changes as a function of environment (the traditional approach), or by measuring spatial correlations (such as mark correlations), that one might discover this correlation.  The question is:  which approach is more efficient, especially for small $\alpha$ when $w_\alpha\approx w + \alpha w\delta$?

\subsection{Measures of environment}\label{sec:enviromeas}
%\textbf{More details about the enviro measures?} 
There are many different methods of quantifying the environment, but most of them can be categorized into those that use a fixed aperture (FA) and others that use near-neighbor (NN) finding.  
A variety of environment measures are analyzed in M12, and we use a subset of these, 
which are briefly described below. 
%$1+\delta_8$ is associated with a fixed spherical? cylindrical? aperture of scale 8$h^{-1}$Mpc, and $\Sigma_N$ which is the nearest neighbour measure with $N=3$.  In both cases, specific choices were made about how to count in the redshift direction (see M12 for details). 
Unless stated otherwise, all of the FA and NN overdensities used in this paper are based on redshift-space distances, as they would be in real data.
In addition, in all cases the density-defining population (DDP) consists of galaxies brighter than the luminosity threshold, $M_r < -19$.

FA measures are often expressed as a local density contrast, 
determined by counting the number of galaxies %of the DDP 
within a given radius, and taking the ratio with the mean density.  
The density contrast is typically defined as 
\begin{equation}
 \delta_g %\equiv \frac{\delta\rho}{\bar\rho}
  = \frac{N_\mathrm{g}-\bar{N_\mathrm{g}}}{\bar{N_\mathrm{g}}} ~,
\end{equation}
where $N_\mathrm{g}$ is the number of galaxies found in the aperture, and 
$\bar{N_\mathrm{g}}$ is the mean number of galaxies that would be expected in 
the aperture if the galaxies were randomly distributed.  
The motivation for using apertures of a particular size is often so that they enclose all of the galaxies within a dark matter halo, while accounting for the effect of redshift-space distortions and redshift uncertainties (Abbas \& Sheth 2005; Gallazzi et al.\ 2009). 
We use $1+\delta_8$, the overdensity in spherical apertures of radius $8~\mathrm{Mpc}/h$ (Croton et al.\ 2005; Abbas \& Sheth 2005).  
Cylindrical apertures and annuli (Gallazzi et al.\ 2009; Wilman et al.\ 2010) yield qualitatively similar results. 

NN measures exploit the fact that objects with nearer neighbors tend to be found in denser environments.  A value of $N$ is chosen that specifies the number of neighbors around the point of interest.  
One can define a projected surface density or a spherical density: 
\begin{eqnarray}
\sigma_{N} &=& \frac{N}{\pi r_{N}^2} \nonumber\\
\Sigma_{N} &=& \frac{N}{(4/3) \pi r_{N}^3} ~,
\label{eq:NN}
\end{eqnarray}
where $r_N$ is the radius to the $N$-th nearest neighbor. 
We use the Baldry et al.\ (2006) measure, which is an average of $\mathrm{log}\sigma_N$ for $N=4$ and 5 with a redshift limit ($\pm\Delta zc=1000\mathrm{km}/\mathrm{s}$) on the DDP, and the  $\Sigma_N$ measure for $N=3$ described in M12. 
In order to use these similarly as the fixed-aperture overdensities, the NN densities need to be normalized, and we do this by defining 
%\begin{equation}
% \delta \equiv \frac{\sigma-\bar\sigma}{\bar\sigma}~,
%\end{equation}
$\delta \equiv (\sigma-\bar\sigma)/\bar\sigma$~,
where $\sigma$ is one of the two density measures described above 
[i.e., $\Sigma_3$ or $(\mathrm{log}\sigma_4+\mathrm{log}\sigma_5)/2$].

\begin{figure}
 \includegraphics[width=\hsize]{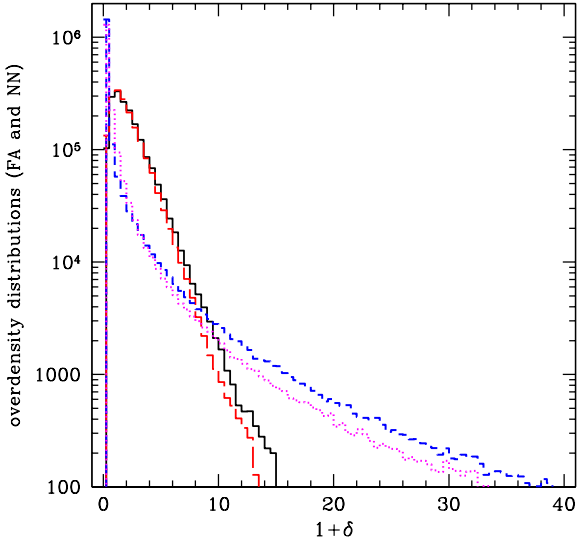}%{mock_overdensity_distsSMz.ps}%dists.ps} 
 \caption{$1+\delta$ overdensity distributions for fixed-aperture (FA)  
          environment measures [$8~\mathrm{Mpc}/h$ spheres 
          (black solid histogram) and cylinders (red long-dashed histogram)], and 
          for nearest-neighbor (NN) environment measures [$\sigma_{4,5}$ 
          (blue short-dashed histogram) and $\Sigma_3$ 
          (magenta dotted histogram)], which have longer tails at large $1+\delta$.}
 \label{fig:deltadists}
\end{figure}

Figure~\ref{fig:deltadists} shows the distribution of these $1+\delta$ overdensities in the mock catalogue.  
Note that the FA overdensity distributions have a similar shape (see also de la Torre et al.\ 2010), as do those of the NN overdensities. 
For simplicity, throughout the rest of this paper we will usually focus on a single FA environment measure, 8~Mpc/$h$ spheres, %which we will refer to as $\delta_8$, 
and a single NN measure, the combination of $\sigma_4$ and $\sigma_5$, which we will henceforth refer to as $\sigma_{4,5}$.

Clearly, the NN measures have longer tails with larger $1+\delta$, consistent with the expectation that they probe smaller scale environments.  (That is, if the NN weights trace the environment within haloes, then we expect them to have $1+\delta\sim 200$; and if the FA weights trace the environment around each halo, they should have $1+\delta\sim 1+\sigma_8$, where $\sigma_8$ is the rms variance of the linear density fluctuation field within $8~\mathrm{Mpc}/h$ spheres.) 
As we show below, mark correlations allow us to quantify this expectation.  But before doing so, we note that standardizing the distributions by subtracting the mean and dividing by the rms still yields distributions with different shapes.

%[segue] 
%The simple quantity $1+\delta$ is commonly used to characterize the overdensity of objects relative to the background density (e.g., Peebles 1980; Mo \& White 1996).  For example, in regions of volume $V$ containing mass $M$, one can define $M/V\equiv{\bar \rho}(1+\delta)$, such that dense regions typically have $\delta>0$ (Cooray \& Sheth 2002; Abbas \& Sheth 2005). The two-point correlation function of the density field, which is the excess probability (relative to a random distribution) of finding a pair of objects at a given separation, can be written as $\xi(\mathbf{r})=\langle\delta(\mathbf{x})\delta(\mathbf{x}+\mathbf{r})\rangle$, where $\delta(\mathbf{x})\equiv(\rho(\mathbf{x})-\bar\rho)/{\bar\rho}$.  The mark two-point correlation function then uses the weighted counts of objects (i.e., a weighted density field), relative to the mean ${\bar w}{\bar \rho}$. 

%\subsection{Removing the Dependence on the Mark Distributions}\label{sec:rescale}

% see markdists_emails.txt
%Mark clustering statistics are sensitive to the mark distributions (Sheth, Connolly \& Skibba 2005; Skibba et al.\ 2009).  
%In order to compare mark correlation functions, we need to account for the shapes of the distributions of the weights, to ensure that only the rank order of the objects matters for the environmental correlations.

%When we apply the additional $1+\delta$ weight, 

%This way, we can be sure that the new weighted and rescaled luminosities (or colors) will only give a different signal if the new weights changed the rank ordering in a way that depends on the environment. 

\subsection{The traditional approach}

%We then sort the new weights and match them to the original sorted weights, thus forcing the new distribution to have the same shape as in the original data set.  (In contrast to the previous section, here we do not use the rank itself as the weight because we want to make a slightly different point.)  Namely, here we explicitly force the distribution of $w'$ to be the same as that of $w=L$ or $w=g-r$.  

In what follows, we will illustrate our results using these FA and NN overdensities. 
%using these FA overdensity $\delta_8$ measured in spheres, as well as the Baldry et al.\ (2006) NN overdensity (using both $\sigma_4$ and $\sigma_5$), described in the previous section. %; using $8h^{-1}$Mpc cylinder overdensities (Gallazzi et al.\ 2009) yields similar results.  
Specifically, in this section we will insert $\delta$ in equations~(\ref{eq:weight}) and~(\ref{eq:rescale}) 
to define $w_\alpha$ for each galaxy, thus adding an environmental dependence to $w$ (luminosity or colour), and then we will measure the distribution of (rescaled) $w_\alpha$ in a number of different overdensity bins.  

\begin{figure}
 \includegraphics[width=0.9\hsize]{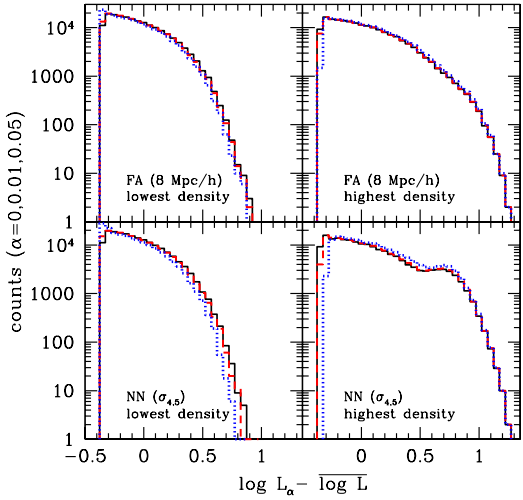}%{lgLnormdists_fixeddensitycompare_10binsnewlabels.ps}%10binslabel.ps}  % for alpha=0,0.01,0.05, try black solid, red dashed, blue dotted 
 \caption{Rescaled $r$-band luminosity functions for bins of environmental overdensity (upper panels: ranked by $8~\mathrm{Mpc}/h$ sphere overdensities; lower panels: ranked by $4^{\rm th}$ and $5^{\rm th}$ NN overdensities).  Only the distributions for the lowest-density (left) and highest-density (right) 10 percent are shown.  
Black solid, red dashed, and blue dotted histograms indicate the distributions for $\alpha=0$, 0.01, and 0.05, respectively.}
 %\caption{{\bf Rescaled} luminosity functions in four bins of environmental overdensity ($8~\mathrm{Mpc}/h$ spheres); those which contain the 10-15, 35-40, 60-65, and 85-90 percentiles.  Black, red, and blue histograms indicate the distributions for $\alpha=0$, 0.01, and 0.05, respectively.}
 \label{fig:LFs}
\end{figure}

Figure~\ref{fig:LFs} shows the results for the luminosity marks. 
The various histograms in each panel show the overdensity dependent luminosity distributions, $p(L_\alpha|\delta)$, for various choices of $\alpha$ (0, 0.01, and 0.05). 
The different panels show different bins in $\delta$ (lowest and highest 10\%; $p(L_\alpha|\delta)$ of intermediate bins have smaller differences), and there are clearly more luminous galaxies and fewer faint galaxies in dense environments. 
The question is whether the differences between the $\alpha=0$ counts and the others are statistically significant.  Obviously, one must be far from $\delta=0$ to see a 
difference; how far is far enough depends on $\alpha$.  Kolmogorov-Smirnov (KS) tests 
suggest that large values of $\alpha$ and/or $\delta$ far from $0$ are required.  Since 
the large $|\delta|$ tails typically contain a small fraction of the full sample 
(Figure~\ref{fig:deltadists}), this technique, in effect, cannot use the vast majority 
of the sample to detect the fact that environment matters.
%By comparing the distributions in different bins, it is evident that there are fewer luminous galaxies in underdense regions than in very dense ones.  More importantly, in any given overdensity bin, the distributions of $p(L|\delta)$ and $p(L'|\delta)$ are clearly very similar, and in fact are nearly identical. (maybe we should quantify this though, e.g., with Kolmogorov-Smirnov probabilities.) 
%We have ranked the galaxies in the catalog by their $8~\mathrm{Mpc}/h$ sphere overdensities, and split them into twenty bins, in 5\% increments.  We have normalized the luminosities $L$ and $L'$ by the mean luminosity of the catalog. 

%Figure~\ref{fig:LFs} shows the luminosity functions for the lowest and highest 10$\%$ overdensities, with $\alpha=0$, 0.01, and 0.5. 
As described earlier in Section~\ref{sec:MCFresults}, to allow for a fair comparison, 
we have rank ordered and rescaled the luminosities so that they have the same  
distributions [$p(>L_\alpha)=p(>L)$, where $L_\alpha$ are the luminosities with $\alpha>0$]. 
%Comparing the distributions in different panels (different overdensity bins), we clearly see that  there are more luminous galaxies and fewer faint galaxies in dense environments. 
Therefore, the overall luminosity distributions are the same by construction, but at fixed overdensity the rescaled ones (with $\alpha>0$) may be shifted from the original one ($\alpha=0$).  
In each panel of the figure, we find that the differences between the luminosity distributions $p(L|\delta)$ and $p(L_\alpha|\delta)$ appear to be very small, except for $\alpha=0$ versus $\alpha=0.05$ with the NN overdensity. 
%[{\bf I can include KS probabilities, but they are usually tiny because $N_{\rm gal}$ is so large.}]

%\textbf{RAS: there are many ways to plot these.  It looks like many environment studies use $\sim10$ overdensity bins, so that's what I used, and I plotted distributions for the lowest/highest overdensities.  The distributions are normalized; not normalizing yields slightly different color distributions.} %(lgL/C)normdists_fixeddensitycompare_10bins.ps
% see calcKSprob.c (uses NR's kstwo).  maybe also consider Anderson-Darling or Kuiper's statistic.
%KS statistics for LF.  FA lo: D=0.01222 for a=0.01, 0.03357 for a=0.05; FA hi: 0.00699, 0.00704; NN lo: 0.01594, 0.06653; NN hi: 0.00821, 0.01752
%       ProbKS for LF.  FA lo: 2.2e-12, 1.2e-90; FA hi: 2.5e-4, 2.1e-4; NN lo: 9.3e-21, 0.0; NN hi: 8.0e-6, 5.3e-25
%KS statistics for C dist.  FA lo: D=0.00182 for a=0.05, 0.00529 for a=0.05; FA hi: 0.00285, 0.003340; NN lo: 0.00759, 0.01477; 0.01550, 0.20864
%       ProbKS for C dist.  FA lo: 0.92, 0.0115; FA hi: 0.44, 1.1e-89; NN lo: 5.0e-5, 6.9e-18; NN hi: 1.2e-19, 0.0

\begin{figure}
 \includegraphics[width=0.9\hsize]{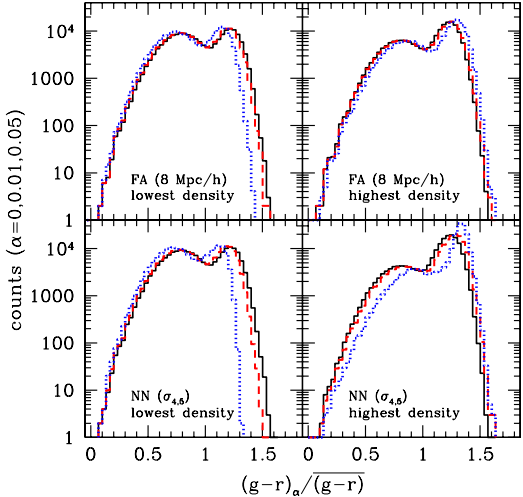}%{Cnormdists_fixeddensitycompare_10binsnewlabels.ps}%10binslabel.ps} 
 \caption{Rescaled $g-r$ colour distributions for bins of environmental overdensity (upper panels: $8~\mathrm{Mpc}/h$ spheres; lower panels: $4^{\rm th}$ and $5^{\rm th}$ nearest neighbors).  Only the distributions for the lowest-density (left) and highest-density (right) 10 percent are shown.  
Black solid, red dashed, and blue dotted histograms indicate the distributions for $\alpha=0$, 0.01, and 0.05, respectively.}
 %\caption{{\bf Rescaled} color distributions for bins of environmental overdensity (ranked by $8~\mathrm{Mpc}/h$ sphere overdensities, in 5\% increments).  Black, red, and blue histograms indicate the distributions for $\alpha=0$, 0.01, and 0.05.}
 \label{fig:colordists}
\end{figure}

Figure~\ref{fig:colordists} shows a similar analysis of (rescaled) $(g-r)_\alpha$ colours. 
Recall that the distributions, when averaged over all $\delta$, have been rescaled to be the same.  
%By comparing the distributions in different bins, 
It is evident that 
there are fewer red galaxies in underdense regions than in very dense ones. 
In addition, in any overdensity bin, the distributions of $p(g-r|\delta)$ and $p[(g-r)_\alpha|\delta]$ are similar, but with significant differences in the red sequence at very low and very high overdensities.  

In general, the weak dependence on $1+\delta$ appears to produce subtle differences in the luminosity and colour distributions.  %$\alpha\ge 0.05$ is required to result in significant differences.
%\textbf{[statistical significance of the different probability distribution functions $p(w)$ and $p(w|\delta)$, where $w$ is luminosity or colour.]} 
We quantify the statistical significance of this with Kolmogorov-Smirnov (KS) tests.  We find that these in fact yield low KS probabilities, indicating that the distributions $p(w|\delta)$, where $w$ is luminosity or colour, do have statistically significant differences, even for low values of $\alpha$.  
However, the significance depends on the number statistics, and typical SDSS galaxy catalogues are at least 25 times smaller than the mock catalogue used here.  
%
%When we account for this, we obtain $P_\mathrm{KS}$=0.44 and 0.06 for $p(L_{\alpha=0.01})$ for the FA and NN overdensities, respectively, and $P_\mathrm{KS}$=0.31 and $3\times10^{-7}$ for $p((g-r)_{0.01})$ for FA and NN overdensities, such that only the latter difference is detectable.  
%At fixed density, $p(L|\delta)$ are not distinguishable for $\alpha=0.01$ but are distinguishable otherwise, while the $p(g-r|\delta)$ distributions have lower $P_\mathrm{KS}$.
When we account for this, at fixed density, we obtain $P_\mathrm{KS}$=0.14 and 0.51 for $p(L_{\alpha=0.01}|\delta)$ in the lowest and highest density bins for the FA overdensities, respectively, making these distributions statistically indistinguishable, while the corresponding colour distributions are marginally distinguishable.  For larger values of $\alpha$, and for the NN overdensities, lower probabilities are obtained ($P_\mathrm{KS}<10^{-3}$), indicating statistically significant differences between the luminosity and colour distributions, especially near the peak of the red sequence. 
%
% normalization issue
Note that for these probability distribution functions, we have normalized by the mean luminosity or colour of the full catalogue; if we normalize by the mean of a given density bin, then $p(w|\delta)$ become more similar ($P_\mathrm{KS}$ close to unity are obtained, indicating virtually identical distributions), except for $\alpha=0.05$ in the highest density bin. %at high density.
%> RAS: The results depend on how the
%> normalization is done though.  Right now I'm normalizing by the mean mark
%> of the full sample (since that's the way the distributions are plotted in
%> Figs. 4 & 5), rather than the mean mark at a fixed density bin.
%RKS: I guess that with PDFs, the crudest thing one might look for is a shift in
%the mean value.  So normalizing by the mean of the sample is probably the
%right thing to do.  If normalizing by the mean at each density bin makes
%things even more indistinguishable, then we can just say so in words.

\begin{figure}
 \includegraphics[width=0.9\hsize]{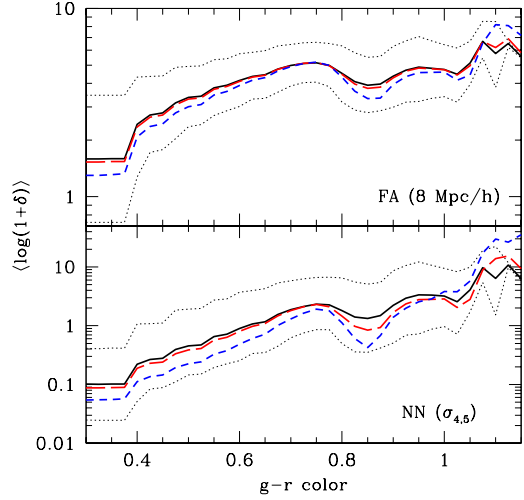}%{mock_Mr19_colordensitycompare_running10b.ps} %{mock_Mr19_colorDCdensitymean_running10.ps} %colorIBdensity
 \caption{The colour-density relation, %of the mock catalogue 
 using $8~\mathrm{Mpc}/h$ sphere overdensities (upper panel) and $\sigma_{4,5}$ nearest-neighbor 
 overdensities (lower panel).  Black, red, and blue lines show the running mean of the relation 
 for $\alpha=0$, 0.01, and 0.05, respectively.  Dotted lines indicate the 1-$\sigma$ range 
 between the 16 and 84 percentiles.} %{\bf RKS:  Do you understand why the curves don't cross?} 
 \label{fig:colordens}
\end{figure}
% to answer Ravi's questions: yes these are rescaled; yes they relations cross but only at red seq.

One can also consider the `colour-density relation' or density-dependent red fraction (e.g., Hogg et al.\ 2003; Balogh et al.\ 2004; Cooper et al.\ 2006; Weinmann et al.\ 2006; Park et al.\ 2007). %and Ball+ 2008; Gruetzbauch+ 2011; Cucciati+ 2012 
% some color-density papers: *Hogg+ 2003; *Cooper+ 2006,2007,2010; *Blanton+ 2005,2006,2007; *Cucciati+ 2006; Baldry+ 2006; *Park+ 2007; *Ball+ 2008; O'Mill+ 2008; SS09; Gallazzi+ 2009; Wilman+ 2010; Ellison+ 2010; Gruetzbauch+ 2011; *Cucciati+ 2012
We show the $g-r$ colour-density relation of the mock catalogue in Figure~\ref{fig:colordens}, using the $8~\mathrm{Mpc}/h$ sphere overdensities.  
%Following Cooper et al.\ (2006), we plot the relation as a function of color rather than the reverse, because the overdensities have larger uncertainties than the colors.  
Note the dip in the relation between the blue cloud and red sequence. 
% mention rescaled color dist.?
The colour-density relations for the colours modified by $(1+\delta)^\alpha$ and rescaled following equation~(\ref{eq:rescale}) are also shown.  
With $\alpha>0$, the colours have been given an additional environmental dependence, and thus we expect them to have a stronger (i.e., steeper) correlation with overdensity compared to $\alpha=0$. 
Evidently, the colour-density relation is only slightly steepened for $\alpha>0$ and would be difficult to detect, depending on the environment measure used (see also M12).  
In addition, the relations only cross for the red-sequence colours, so the densities in overdense regions must be accurate in order to detect different environmental trends.

%[\textbf{is this useful?}  if so, I could add contours to the plot.  or we could show $f_\mathrm{red}(1+\delta)$ instead.]  %would it be better to move it to the end as an appendix?] 

% could also show mock_Mr19_lgLdensitycompare_indepbins.ps here, but I don't think it's necessary.
The analogous `luminosity-density' relation (not shown) has a similar shape as the 
luminosity-halo mass relation (e.g., More et al.\ 2009), except that at luminosities fainter 
than $L_\ast$ (the break in the luminosity function), the relation has increased scatter 
and is no longer monotonic.  This behavior is not due to faint satellite galaxies in group/cluster environments, which are outnumbered by faint `field' galaxies; it is instead due to the fact that both FA and NN environment measures do not accurately probe the environments of low-mass haloes (see M12 for details).
%Ravi: Since it is true that non-central galaxies in clusters can have small L, one does expect a large scatter in environments at small L.  We should say explicitly why this is not the dominant effect at small L (I guess there are many more faint `field' galaxies than faint non-central cluster galaxies?).  Ramin: I think you're right that even though there are faint non-central galaxies in clusters, they are outnumbered by faint 'field' galaxies.

\subsection{Mark correlations}
\begin{figure*}
 \includegraphics[width=0.497\hsize]{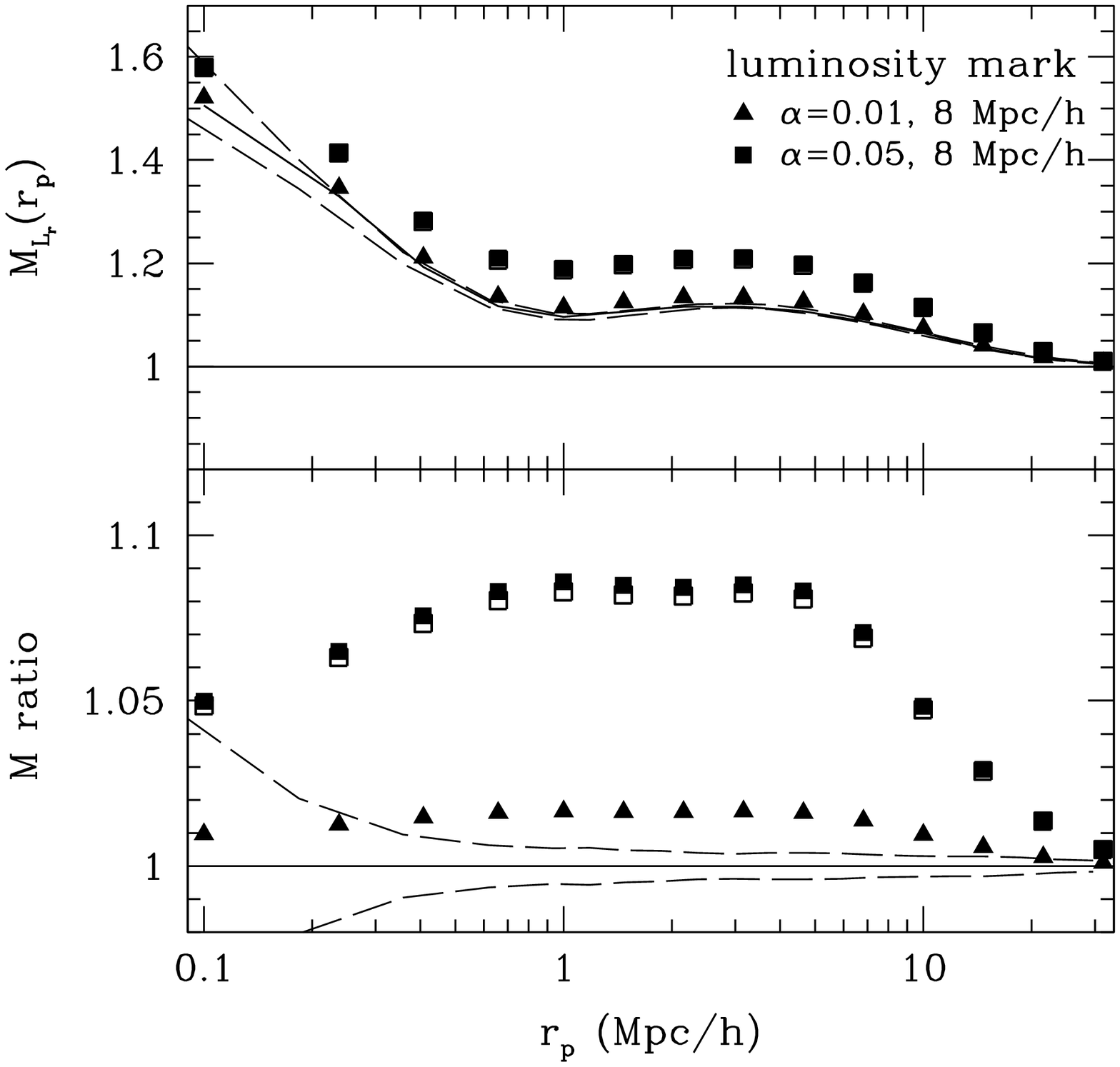} %{mock_Mr19_projnew_Lmark_DCenviro8Mpc_2alpha_errors_labels.ps}
 \includegraphics[width=0.497\hsize]{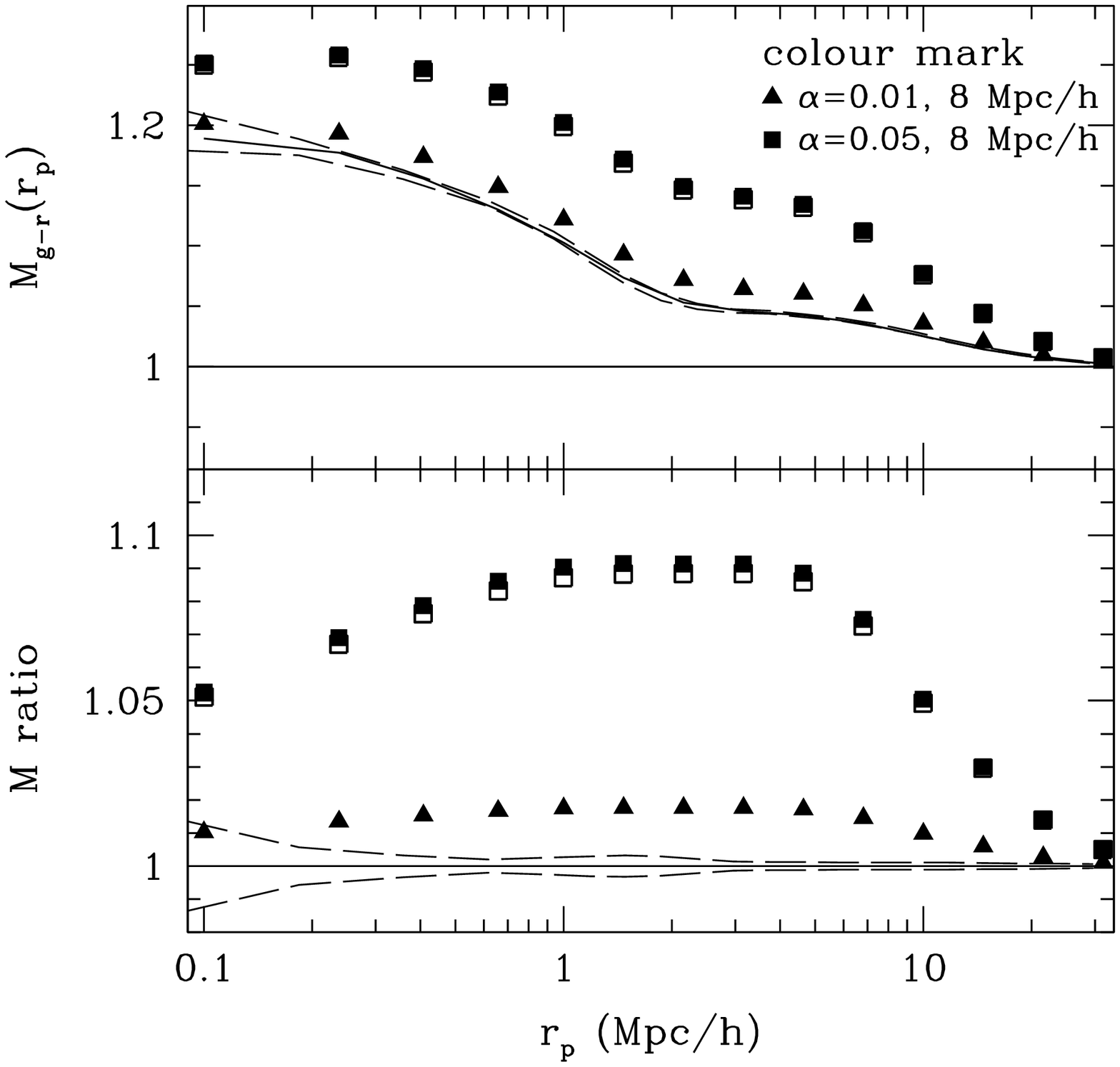} %{mock_Mr19_projnew_Cmark_DCenviro8Mpc_2alpha_errors_labels.ps}
 \caption{$r$-band luminosity (left) and $g-r$ colour (right) mark correlation functions (upper panels), with additional environmental correlations using FA overdensities:  $(1+\delta)^\alpha$ with $\alpha=0.01$ (red triangles) and 0.05 (blue squares). 
          The lower panels show the ratio of $M(\alpha\ne0)/M(\alpha=0)$, to more clearly indicate the effect of the additional environmental correlations. 
          The $\alpha$-dependence of the effect on the mark signal can be easily estimated (open squares; see Appendix~\ref{app:2alpha}). The dashed lines show the uncertainty of the measurements. %, while the short-dashed lines show the uncertainty of similar measurements using real data (see Fig.~\ref{fig:mockvsdataMCFs}.)
        }
 \label{fig:DCenviro}
\end{figure*}

%Having shown that it is difficult to detect the environmental correlation using the traditional approach, we now turn to the mark correlation measurement.  
We now turn to the mark correlation measurements, as an alternative to the traditional approach to quantifying environmental correlations. 
%\ref{fig:DCenviro} and \ref{fig:IBSMenviros}
The result is shown in Figure~\ref{fig:DCenviro}, for $\alpha=0.01$ and 0.05, using the FA overdensities.  

Note that even a dependence as weak as $(1+\delta)^{0.01}$ results in a significantly stronger signal, while the effect of $\alpha=0.05$ is substantially larger still. To clearly demonstrate this, the lower panels of the figures show the ratio of these mark correlation functions (triangle and square points) to that of the unmodified ($\alpha=0$) mark correlation functions (solid curves).  
The dashed lines show the jack-knife errors\footnote{Statistical errors are estimated with ``jack-knife" resampling, using 27 subsamples of the full mock cube. 
%and 30 subsamples of the SDSS catalogue. 
The variance of the clustering measurements yield the error estimates.  (For details, see Zehavi et al.\ 2005; Norberg et al.\ 2009.)} 
of the clustering measurements, %for the mock and SDSS catalogues. 
indicating that the environmental correlations with $\alpha=0.01$ are detectable except for the smallest separations.  
%
%For comparison, the short-dashed lines show the errors of similar measurements in SDSS; these errors are larger, but but they nonetheless indicate that a dependence on $(1+\delta)^{0.01}$ is detectable at least for colour marks.  
%\textbf{[revise this.]}
We have also estimated jack-knife errors of similar measurements of the SDSS catalogue described in Section~\ref{sec:sdss}, and have found that these are systematically larger than the errors of the mock catalogue's mark correlations.  Nonetheless, we find that a dependence on $(1+\delta)^{0.01}$ is still detectable at least for colour marks.  

%In an appendix, we show that the effect on WW/DD is to introduce an additional environmental correlation that is approximately proportional to $2\alpha$.  %This is a manifestation of the fact that 
Note that the quantitative effect of $\alpha$ on the mark correlation functions is similar.  In Appendix~\ref{app:2alpha}, we argue that the $(1+\delta)^\alpha$ weighting adds a new mark signal which, for $\alpha\ll 1$, is proportional to $2\alpha$.  Therefore, given the mark correlation with a particular value of $\alpha$ dependence (e.g., $\alpha=0.01$), we can predict the signal for a different $\alpha$ (e.g., $\alpha=0.05$).  This expectation is borne out: the predicted mark correlation functions with $\alpha=0.05$ are nearly identical to the measured ones.

\begin{figure*}
 \includegraphics[width=0.497\hsize]{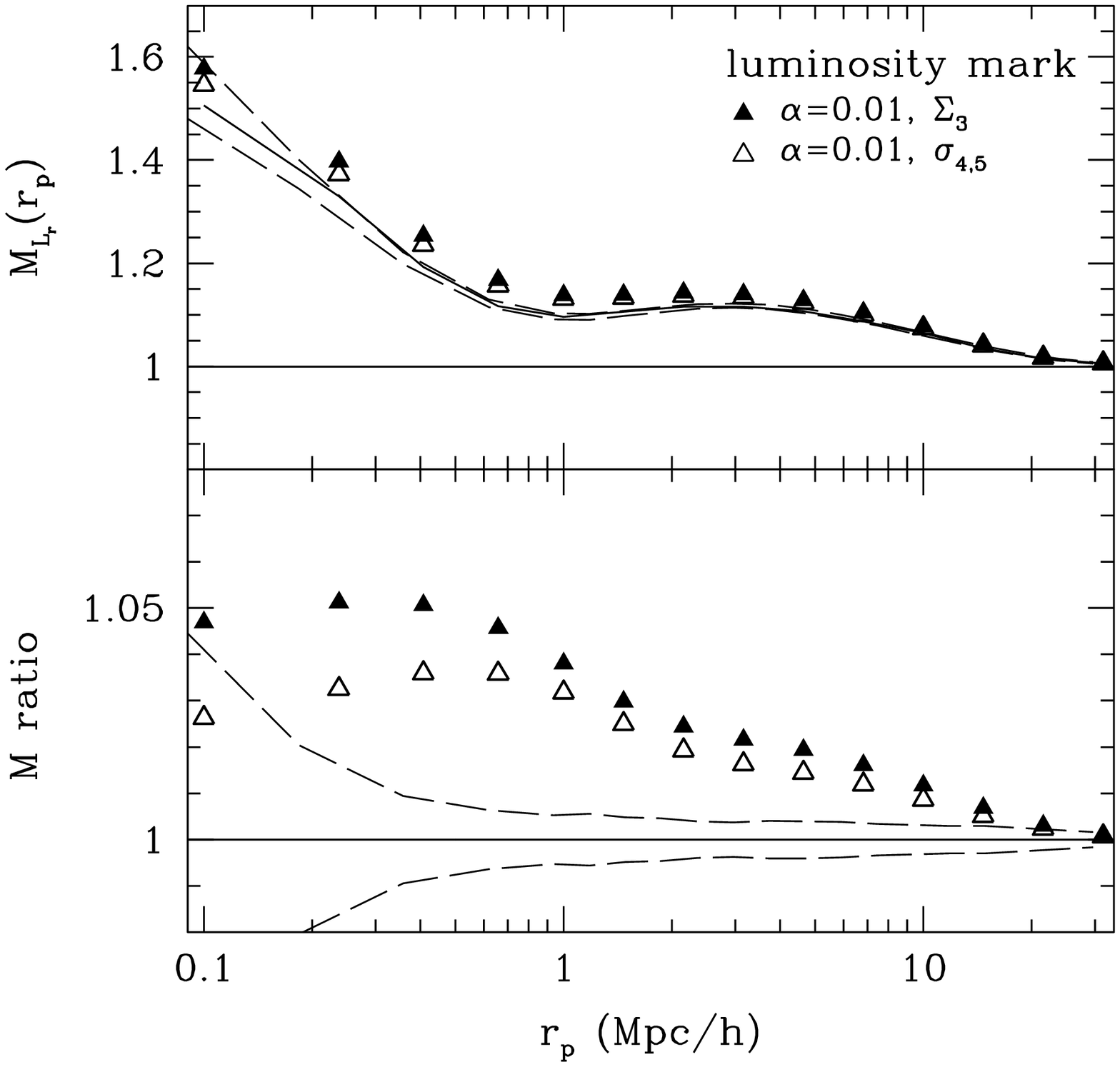} %{mock_Mr19_projnew_Lmark_IBandSMzenviros_errors_labels.ps}
 \includegraphics[width=0.497\hsize]{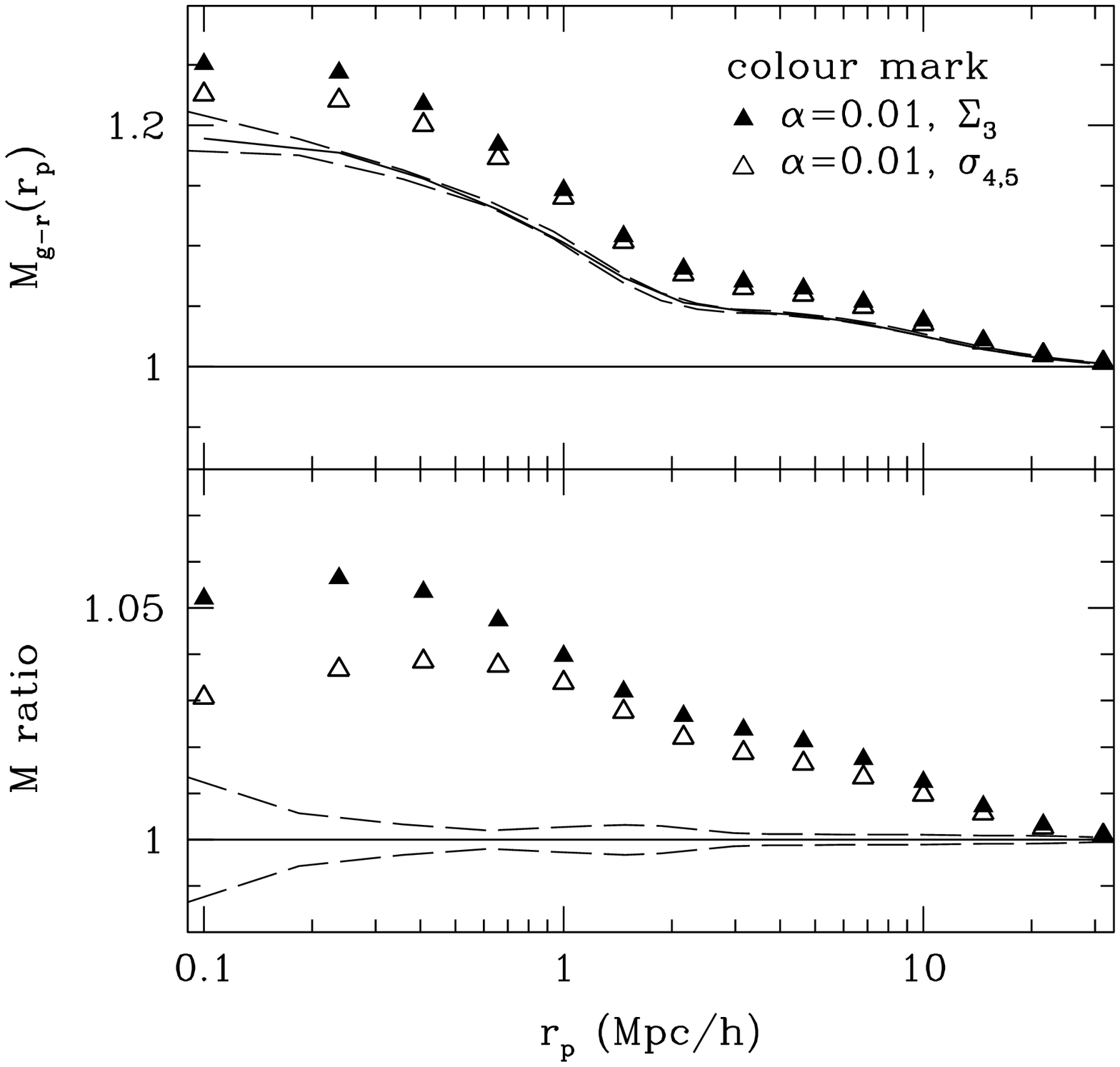} %{mock_Mr19_projnew_Cmark_IBandSMzenviros_errors_labels.ps}
 \caption{$r$-band luminosity (left) and $g-r$ colour (right) mark correlation functions (upper panels), like Figure~\ref{fig:DCenviro}, but with additional environmental correlations using NN overdensities: third nearest neighbor (open triangles) and fourth and fifth nearest neighbors (solid triangles). 
The results are shown for $\alpha=0.01$, and the effect is relatively strong because of the long tail in the NN overdensity distributions (Fig.~\ref{fig:deltadists}).}
 \label{fig:IBSMenviros}
\end{figure*}

We show analogous mark correlation functions using NN overdensities in Figure~\ref{fig:IBSMenviros}.  These overdensities are clearly more sensitive to small-scale environmental correlations than large-scale ones. 
%and this is especially the case when fewer neighbors are used (open triangles).
In addition, the new mark signal for $\alpha=0.01$ is larger than for the FA overdensities (triangle points in Fig.~\ref{fig:DCenviro}), which is due to the long tail of the NN overdensities at large $1+\delta$.  In Section~\ref{sec:markstats}, we account for this effect and find that these overdensities are particularly sensitive to environmental correlations on scales smaller than $600~\mathrm{kpc}/h$. 
An advantage of the mark correlation approach is that it quantifies the scale dependence of environmental trends, and that it exploits the entire dataset (rather than splitting a sample with overdensity cuts, for example).

%discuss dependence on aperture size \& annuli (probing different scale enviros), and NN number.  % e.g., MCFs w/ small scale enviros (Anna's 2 Mpc aperture, Wilman \& Zibetti's s-scale annuli.
%Note that when fewer neighbors are used, the mark correlations are more enhanced at small separations, indicating that these overdensities are probing especially small-scale environments. Similarly, for the fixed-aperture overdensities, it is not surprising that when smaller apertures (or annuli, using Wilman et al.\ (2010) overdensities) are used, smaller-scale environmental correlations are more evident. 
%consistent with the results in M12.

%\section{Results: Distributions of Galaxy Properties}\label{sec:dists}

%\textbf{which plots to show?} 

%In this section we will show luminosity and color distributions at fixed overdensity, and assess to what extent they vary when we add the extra environmental dependence.  That is, we will show distributions of $p(w|\delta)$ and $p(w'|\delta)$, where $w$ is the weight (luminosity or color) and $w'$ is the weight modified by $(1+\delta)^\alpha$. 

% maybe cite Muldrew and other papers, like de la Torre et al. (2010)

\section{Rank-ordered mark correlations}\label{sec:markstats}

%In Section~\ref{sec:markstats} we show how to remove the effect of the mark correlation signal on `units', arguing that for any given weight, one should simply rank order and use the rank as the weight.  
%In Section~\ref{sec:markstats}, we account for this effect and find that these overdensities are particularly sensitive to environmental correlations on scales smaller than $600~\mathrm{kpc}/h$.

\subsection{The traditional WW/DD measurement}
We now demonstrate how mark correlation functions quantify the scale dependence 
of environmental correlations by using the FA and NN estimates of the local density around galaxies, $1+\delta$, as the weight (or mark). 
As these quantities are intended to be direct probes of galaxy environments, rather than 
indirect ones such as luminosity or colour, one would expect stronger mark correlation signals 
than those obtained in the previous section.

%Figure~\ref{fig:densityMCFs} shows WW/DD for galaxies with $M_r<-19$ 
%in the mock catalogue where the FA and NN estimates of the local density 
%around each galaxy were used as weights. 
%Filled symbols show the original measurement; 
%the NN weights produce a much stronger signal, especially on small scales. 
The filled symbols in Figure~\ref{fig:densityMCFs} show the result. 
The NN weights, using $\sigma_{4,5}$ overdensities, produce a much stronger signal, 
especially on small scales.  
This is consistent with previous work, in which a NN local density was used as a weight (White \& Padmanabhan 2009).  
In Appendix~\ref{app:envwt}, we also illustrate %with a model 
the effect of using the small-scale environment as a weight. 

The obvious jump in amplitude at $r\le 2h^{-1}$Mpc for the NN weight 
is consistent with the expectation that it probes scales within haloes.  
In addition, the fact that the FA signal reaches a maximum at $1h^{-1}$Mpc, 
which is roughly the scale of a group or cluster, suggests that these are the 
pairs which are in the densest larger-scale environments. 
The decrease on smaller scales indicates that an increasing fraction of 
the closest pairs, which may be low-mass interacting galaxies, are not  
in particularly dramatic larger-scale ($\sim 8h^{-1}$Mpc) overdensities.  
%\textbf{[cite M12 \& Haas+ 2012 here?]}
% maybe also mention Shattow \& Croton, in prep.?

On the other hand, comparing the FA signal to the NN signal is less 
straightforward.  For example, it is not clear what to make of the fact 
that the two weights have the same amplitude at scale $r\ge 10h^{-1}$Mpc 
(other than that pairs separated by $10h^{-1}$Mpc have weights which 
are above average by the same factor).  This is because the two sets 
of weights have rather different distributions 
(Figure~\ref{fig:deltadists}).  We explore how to remove this in the 
next section.  

\subsection{Rank ordered marks}\label{sec:rankorder}
\begin{figure}
 \includegraphics[width=\hsize]{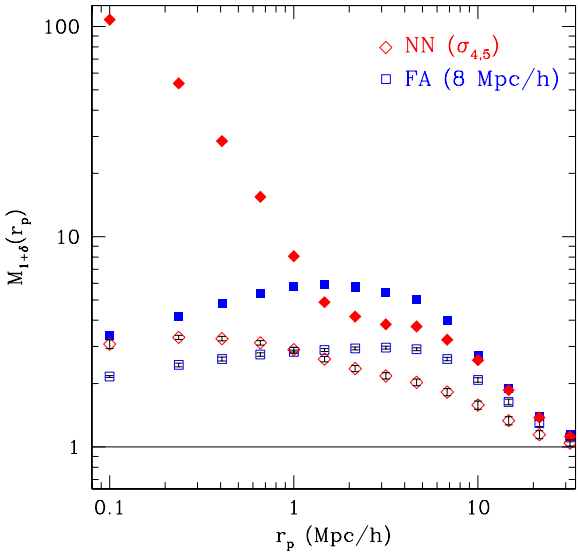} %{mock_Mr19_projnew_1plusdmark_lgMp_rescalingeffect_errors_v2.ps}
 \caption{Mark correlation functions %for galaxies with $M_r<-19$ in the SDSS DR7 
          where FA and NN estimates  
          estimates of the local density (8~Mpc/$h$ spheres and $\sigma_{4,5}$, respectively) were used as marks.  
          Filled symbols show the original measurement, and open 
          symbols show the result of rank ordering and rescaling 
          to [0,1] before making the measurement.  
          %Error bars show jack-knife errors on the measurements, 
          %and dotted lines show the scatter from randomly scrambling 
          %the marks (see text for details). 
    }
 \label{fig:densityMCFs}
\end{figure}

As stated previously, the strength of mark correlations is affected by the 
shape of the marks' distributions, which makes it difficult to fairly compare 
the mark correlations of different marks.  
To remove the effect of the distributions, we first rank order the marks.  
We could then have scaled one of the distributions to the 
other, but this would not allow us to compare either of these with 
a third mark, for example.  
%This can be done, once and for all, simply by 
Instead, as a more general solution, we perform the  
rank ordering and then \textit{use the rank itself as the mark}.  
The open symbols in Figure~\ref{fig:densityMCFs} 
show the result of doing this and then remeasuring WW/DD.  
(In practice, we rank order and then match to a uniform distribution 
on [1,N].  In this way, all marks are scaled to the same %(uniform) 
distribution, so the mark correlation signal can be compared between 
marks.  However, the matching to a uniform distribution is not really 
necessary.)

The rescaling changes the ratio of the small- to large-scale signal 
dramatically, particularly for the NN weights, for which the required 
rescaling is much larger.  Evidently, the large $\delta$ tail in 
Figure~\ref{fig:deltadists} contributes significantly to the small $r$ 
signal; while not unexpected, it is nice to see this confirmation 
that the NN weights really do correspond to small scale environments.  
(We will return to the flatness of the signal shortly.) 

On the other hand, notice that now the feature at $1h^{-1}$Mpc in the 
FA signal has gone away.  This shows that rescaling comes with a 
cost, since there may be information in the shape of the distribution, 
which rank-ordering removes.  
%
% the analytic rank ordered calculation of reds and blues has been moved to an appendix.
To illustrate the effect of rank ordering, a model based on cluster and field populations is described in Appendix~\ref{app:clusterfield}.

\subsection{Effect of redshift-space distortions}

\begin{figure*}
 \includegraphics[width=0.497\hsize]{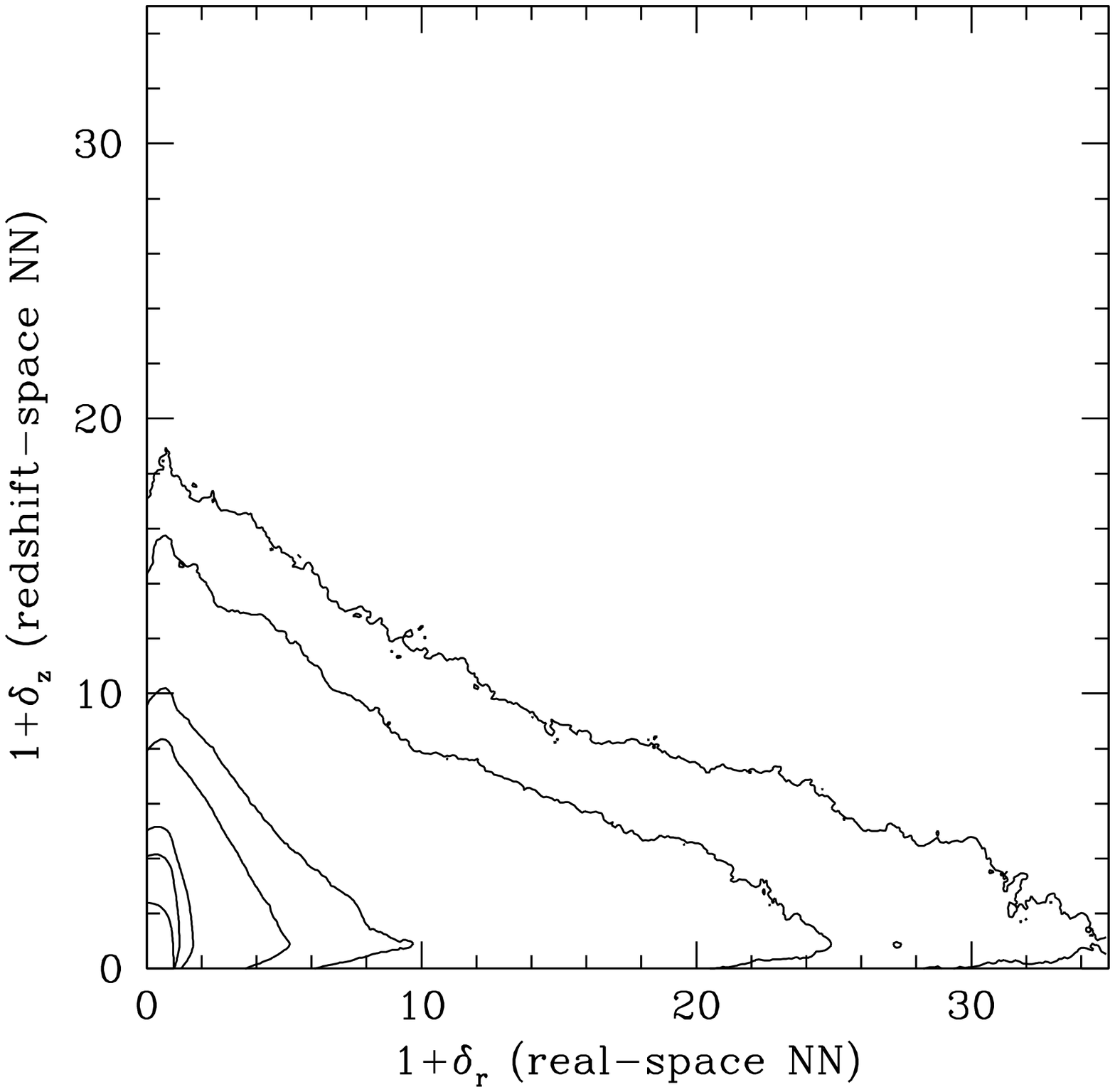}%{mock_PCoverdensities_NNrvsNNzcontours.ps}
 \includegraphics[width=0.497\hsize]{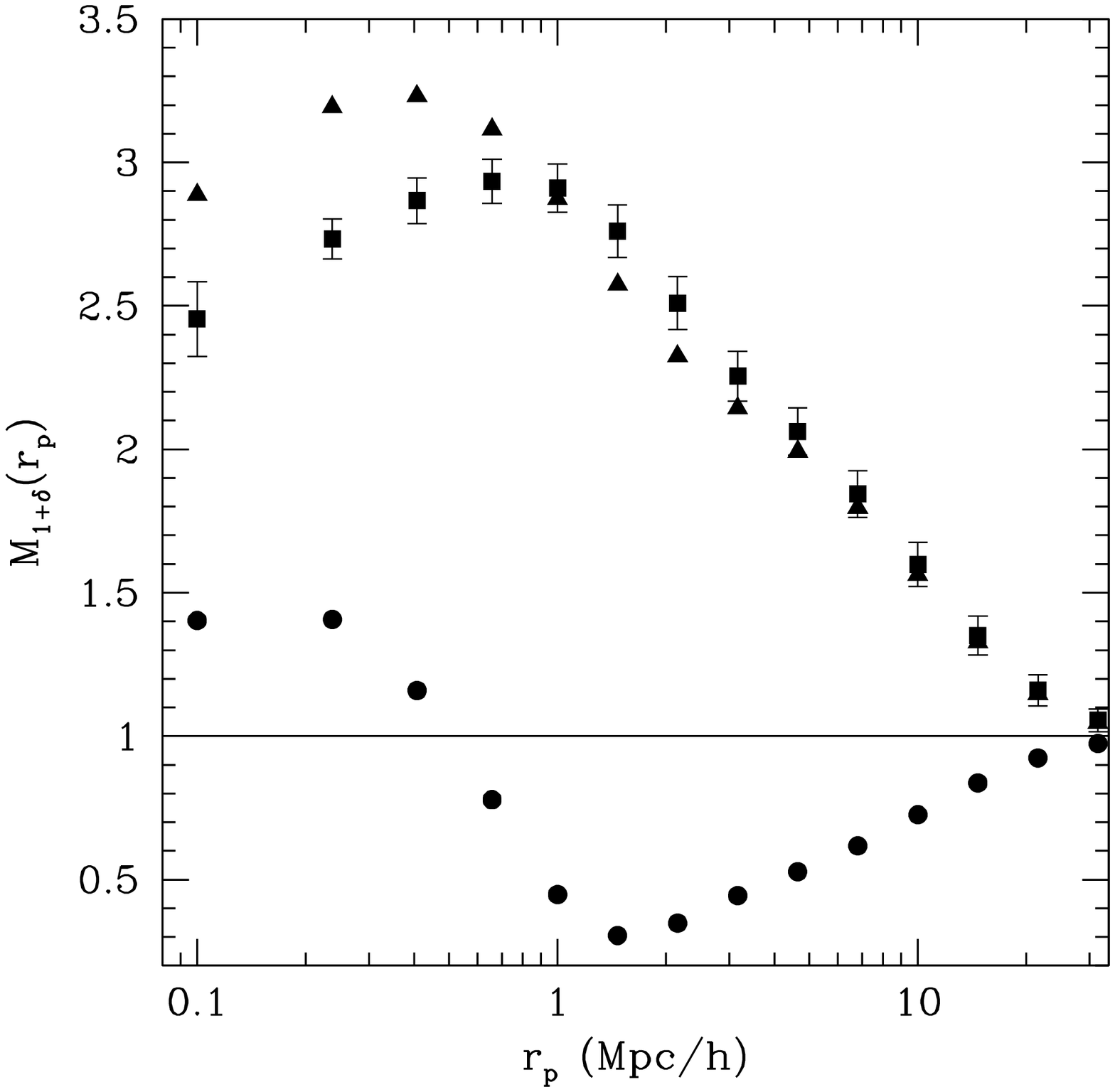}%{mock_Mp_1plusdmark_NNrvsNNz_v2.ps} %v1c.ps
 \caption{Left: contour plot of real-space vs $z$-space Park et al.\ (2007) NN overdensities. 
          Right: rank-ordered and rescaled mark correlation functions of these real-space and $z$-space overdensities (triangle and square points, respectively), analogous to the mark correlations in Fig.~\ref{fig:densityMCFs}. %, compared to that of Baldry et al.\ (repeated from Fig.~\ref{fig:densityMCFs}).  
          The $\delta_r-\delta_z$ mark correlations (circles) are also shown.}
 \label{fig:NNrNNz}
\end{figure*}

% see e-mail about NNr vs NNz overdensities.
% mock_PCoverdensity_dists.ps, mock_PCoverdensities_NNrvsNNzcontours.ps, mock_Mr19_PCdeltadiff_dist.ps, mock_Mp_1plusdmark_NNrvsNNz.ps
%\textbf{[discuss effects of finger-of-god $z$-space distortions, and show figure~\ref{fig:NNrNNz}.]}  %include the $\delta_r-\delta_z$ too?
%[better segue, motivation.] %and mention choice of Park et al. overdensities?
Throughout this paper we have used environment measures based on redshift-space distances.  
It is important to consider the effect of redshift-space distortions on these environments, in the context of rank-ordered mark correlations. 
%In addition to the above, it is important to consider redshift-space distortions in this context.  
The nonlinear virial motions of galaxies in haloes spread out objects in groups and clusters along the line-of-sight to produce `fingers-of-god' (FOG; Jackson 1972; Peebles 1980). 
These small-scale distortions can affect how galaxy environments are assessed, as we see using the Park et al.\ (2007) NN environment 
measures\footnote{Park et al.\ (2007) local densities are estimated by using 20 nearest neighbor galaxies, with the galaxies centrally weighted by a spherical adaptive smoothing kernel (and hence not equivalent to $\Sigma_{20}$ as defined in Eqn.~\ref{eq:NN}).} 
in Figure~\ref{fig:NNrNNz}. 
In the left figure, the real-space and $z$-space overdensities clearly have considerable scatter 
between them, and in $z$-space there is a deficit of very large overdensities, due to the FOG spreading them out.  

This is also seen in the mark correlation functions (right panel, analogous to the rescaled mark correlations in Fig.~\ref{fig:densityMCFs}).  
The FOG distortions can result in underestimated densities for small-separation pairs $r_p<1~{\rm Mpc}/h$, but overestimated densities for more widely separated pairs ($r_p>1~{\rm Mpc}/h$) where the FOG reach into underdense regions (see also Abbas \& Sheth 2007). 
We have tested this using smaller-scale overdensities ($\Sigma_3$, used in Fig.~\ref{fig:IBSMenviros}), which have a similar result but the transition between underestimated and overestimated overdensities occurs at smaller separations.  
%[explain small-scale downturn.]
Note that the small-scale downturn at $r_p<400~{\rm kpc}/h$ in the mark correlations (in the right panel of Fig.~\ref{fig:NNrNNz} and in Fig.~\ref{fig:densityMCFs} with the FA overdensity marks) is \textit{not} due to FOG, since it occurs with the real-space overdensities as well; it occurs because these environment measures best probe larger-scale environments, as opposed to environment measures with smaller apertures or fewer neighbours. 
%As stated in Section 3.1, the environment measures used throughout the paper use redshift-space distances, except for the results shown in Figure 10.  The small-scale downturn of the overdensity mark correlation (and of the FA overdensity mark correlation in Fig. 9) occurs because the scales that are best probed by these measures are at larger scales (0.5-1Mpc in Fig. 10), and smaller-scale environments (with few galaxies in small, low-mass haloes) are not probed as well.   The small-scale downturn is then not only due to the FOG distortions.

Note too that even though the $\delta_r-\delta_z$ mark correlations depart from unity at both small and large scales, rescaling to a uniform distribution nonetheless results in $1+\delta$ mark correlations that are approximately consistent 
%with the previous Baldry et al.\ measurement (repeated from Fig.~\ref{fig:densityMCFs}), 
with the real-space ones especially at larger separations, 
demonstrating the utility of the rank-ordered mark correlations.

\subsection{Colour and luminosity}\label{sec:LCrankMCFs}
\begin{figure}
 \includegraphics[width=\hsize]{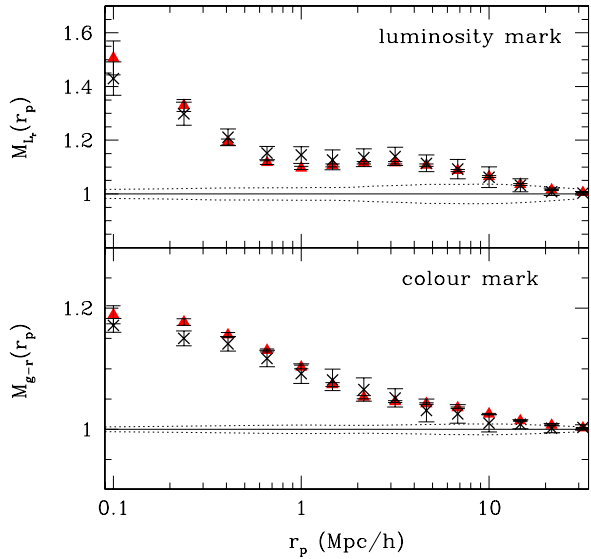}%{mockvssdss_Mr19_LCmarkCFs_v2c.ps}
 \caption{$r$-band luminosity and $g-r$ colour mark correlation functions 
	  for galaxies with $M_r<-19$ in the SDSS DR7 (crosses) and in 
	  the mock catalogue (triangles).
 %   Mark correlation functions are the ratio of the weighted and unweighted correlation functions (upper panel of Fig.~\ref{fig:2PCF}). 
	  Error bars show jack-knife errors on the measurements, 
	  and dotted lines show the scatter from randomly scrambling 
	  the marks (see text for details). 
    %\textbf{show these and/or MCFs w/uniform mark dists.?}
    }
 \label{fig:mockvsdataMCFs}
\end{figure}

We remarked in the introduction that it is difficult to compare the usual measurements of the colour and luminosity mark correlations with one another.  These measurements are shown in Figure~\ref{fig:mockvsdataMCFs}.  Note that, in contrast to the previous section, here we also compare measurements in the mocks with similar measurements in the SDSS.  The agreement between the triangles and crosses shows
that the mock catalogues faithfully reproduce the luminosity and colour dependence of clustering; since these WW/DD signals were not used to construct the mocks, they represent nontrivial tests of the mock-making algorithm.  

\begin{figure}
 \includegraphics[width=\hsize]{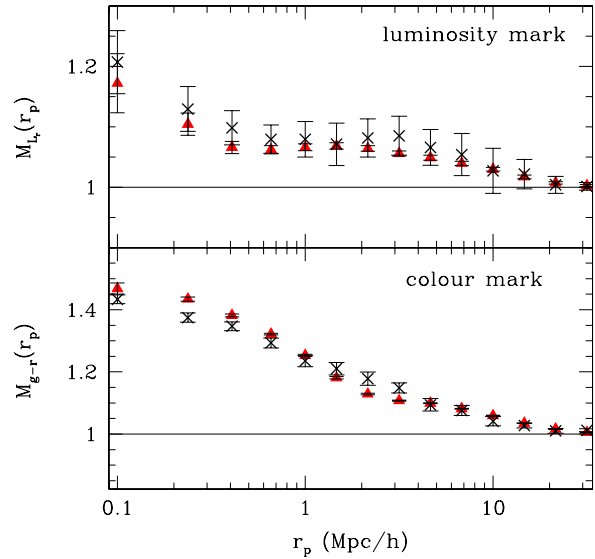} %{mockvssdss_Mr19_LCmarkCFs_uniformdists2c.ps}
 \caption{Same as previous figure, but now the weights have all been 
          rank-ordered and then scaled to a uniform distribution.  
          Although the qualitative trends are the same as in the previous 
          Figure, now the signal is stronger in the bottom panel 
          (i.e., when weighting by colour), indicating that the 
          colour-density correlation is stronger than is 
          luminosity-density. }
 \label{fig:unimarkCFs}
\end{figure}

While this is reassuring, a puzzle lies in the fact that we naively expect colour to correlate more strongly with environment than luminosity 
(e.g., Butcher \& Oemler 1984; Bower et al.\ 1998; Diaferio et al.\ 1999; Blanton et al.\ 2005), %maybe discuss this further...
% Bower R. G., Kodama T., Terlevich A., 1998, MNRAS, 299, 1193
% Diaferio A., Kauffmann G., Colberg J. M., White S. D. M., 1999, MNRAS, 307, 537
but the WW/DD signals do not show this: the amplitude of the luminosity mark correlations is stronger than that of colour. 
This is primarily because the two weights have very different distributions (Figs.~\ref{fig:LFs} and \ref{fig:colordists}); that of luminosity is much broader, such that bright galaxies have $L\gg{\bar L}$. 
% for L, $mmax/$mmean=86.6, for C, $mmax/$mmean=1.62 
Figure~\ref{fig:unimarkCFs} shows the result of rank ordering and using the rank as a mark instead.  Unlike the previous figure, now colour clearly produces the stronger signal.  
% could also mention stellar mass here (Bell+; Zibetti+), which produces an even stronger signal, since it's related to both luminosity and color
It is also worth noting that the error bars are much larger for the $L$-weighted signal, showing that a large range of $L$-ranks contributes at each $r$; this range is much narrower for $g-r$ colour.  %{\bf RKS:  So I guess it would be nice to see the error bars on the FA and NN measurements.}  

Finally, our rank ordering procedure allows us to compare these measurements with those in Section~\ref{sec:rankorder}. 
Comparing the luminosity and colour mark correlation functions (Fig.~\ref{fig:mockvsdataMCFs}) 
to the local-density mark correlations (Fig.~\ref{fig:densityMCFs}) 
shows that both luminosity and colour produce significantly weaker signals.

%\subsection{Description}\label{sec:MCFs}

%[also look at description in SS09, Sheth05.]  %(the following is adapted from S+12.)
%[this is similar to the description in other papers.  should we shorten it?]

%We characterize galaxies by their properties, or ``marks", such as their luminosity, colour, morphological type, stellar mass, star formation rate, etc. In most galaxy clustering analyses, a galaxy catalogue is cut into subsamples based on the mark, and the two-point clustering in the subsample is studied by treating each galaxy in it equally (e.g., Norberg et al.\ 2002; Madgwick et al.\ 2003; Zehavi et al.\ 2005; Coil et al.\ 2006).  %Li et al. 2006; Tinker et al.\ 2008.  These studies have shown that galaxy properties are correlated with the environment, such that elliptical, luminous, and redder galaxies tend to be more strongly clustered than spiral, fainter, and bluer galaxies.

%Nonetheless, the galaxy marks in these studies are used to define the subsamples for the analyses, but are not considered further.  This procedure is not ideal because the choice of critical threshold for dividing galaxy catalogues is somewhat arbitrary, and because throwing away the actual value of the mark represents a loss of information.  In the current era of large galaxy surveys, one can now measure not only galaxy clustering as a function of their properties, but the spatial correlations of the galaxy properties themselves.  We do this with ``marked statistics", in which we weight each galaxy by a particular mark, rather than simply count galaxies as ``one" or ``zero".

\section{Conclusions and Discussion}\label{sec:conc}

Our key results can be summarised as follows: 

\begin{itemize}
  \item Mark correlation functions are particularly sensitive to environmental correlations, specifically when using enhanced weights of $(1+\delta)^\alpha$ with $\alpha>0.01$, though this sensitivity depends on the environment measure, scale, and the mark's uncertainty.
  \item Small environmental correlations are difficult to detect with more traditional methods, as highlighted by the (lack of) variation of most of the mark distributions at fixed overdensity, and in the colour-density relation.  
  \item Rank ordering the marks and then using the rank as the weight provides a simple way to compare results for different marks, because it removes any dependence on the marks' distributions. 
  %\item other conclusions?
\end{itemize}

%discuss utility of mark clustering statistics vis-\'{a}-vis other environment measures: 
%advantages and disadvantages...
The analysis in this paper highlights the advantages (and disadvantages) of mark clustering statistics. 
The fact that they are sensitive to weak environmental correlations that traditional methods have difficulty detecting demonstrates their utility. 
Mark statistics are particularly useful for identifying and quantifying environmental trends. 
This owes to the fact that the statistics of entire samples can be folded together, 
%rather than binning them or attempting to split them into `field' and `cluster' subsamples, for example. 
producing clearer correlations than simply binning environments or splitting galaxies into `field' and `cluster' subsamples, for example. 
Nonetheless, one cannot determine from these trends alone which galaxies occupy which environments; more information is needed (e.g., from halo models of galaxy clustering) in order to associate particular galaxies with environments of particular halo mass or overdensity (Skibba et al.\ 2006; Skibba \& Sheth 2009). %and Sheth 2005?

In contrast, methods that characterize individual local galaxy environments can associate galaxies with overdensities, though they too have strengths and weaknesses.
Fixed-aperture and nearest neighbor overdensities are sensitive to inter- and intra-halo environments, respectively, consistent with the findings of M12 and Haas et al.\ (2012). 
We showed this with the scale-dependent mark clustering measurements, which overlapped at 
$r_p\sim600~\mathrm{kpc}/h$, within the `one-halo term'; fixed apertures, if sufficiently large, can encompass entire haloes as well as some of the surrounding regions. %while nearest neighbors...
Nonetheless, the interpretation of environmental trends can be difficult, and depends crucially on how the overdensities are measured and on the density-defining population.

One can also interpret our results in terms of central and satellite galaxies in haloes. %(e.g., Berlind et al.\ 2003; Kravtsov et al.\ 2004). 
%note though that the distinction between them breaks down in massive haloes 
%(e.g., Skibba et al.\ 2011). 
Since satellite luminosities, colours, and stellar masses depend only weakly on halo mass, 
and hence only weakly on the environment 
(e.g., Skibba et al.\ 2007; van den Bosch et al.\ 2008; Skibba 2009; see also Neistein et al.\ 2011; De Lucia et al.\ 2012), 
the majority of the environmental correlations that we detect are due to the central galaxies. 
%For example, unlike in low-mass haloes, centrals in massive haloes are almost entirely on the red sequence; since these haloes are easily identifiable as overdense environments, the color mark correlations with $\alpha>0$ are strongly enhanced by both fixed-aperture and NN environment measures.
For example, the colours of central galaxies are strongly halo mass dependent, and this is clearly shown by the colour mark correlations, which are especially sensitive to the dependence on overdensity. 

%maybe say something about trends with mean galaxy properties vs distributions of the properties.

%Darren: A bit of a sudden end. Any (not too profound) closing statement we can make, perhaps about future directions or applications?
Finally, we note that
%mark clustering statistics, and rank-ordered mark correlations in particular, 
rank-ordered mark correlation functions are applicable to any 
comparative analysis of environmental trends involving large catalogues of objects in surveys or simulations with sufficiently accurate distances and marks, and are useful for testing or constraining models. %(Sheth, Connolly \& Skibba 2005; Skibba \& Sheth 2009; Padmanabhan \& White 2009).
%maybe also mention assembly bias?  could cite Croft et al. (arXiv:1109.4169) here or with White-Pad.  other relevant papers: Zhu et al. (2006, 6ApJ, 639, L5), Croton et al. (2007), Harker et al. and Wechsler et al.
% (e.g., Wechsler et al.\ 2006; Harker et al.\ 2006; Croton et al.\ 2007; Croft et al.\ 2011).
Rank-ordered mark correlations could be useful for quantifying and comparing measures of `halo assembly bias' (e.g., Sheth \& Tormen 2004; Wechsler et al.\ 2006; Harker et al.\ 2006; Croton, Gao \& White 2007; Croft et al.\ 2012), such that halo formation time, concentration, or occupation is weakly correlated with the environment at fixed mass. 
These statistics could also be applied to tests of `halo abundance matching' (HAM; Vale \& Ostriker 2006; Conroy, Wechsler \& Kravtsov 2006; Neistein et al.\ 2011; Trujillo-Gomez et al.\ 2011; Kang et al.\ 2012) %also Simha et al. (2012) and maybe Kang et al. 
methods, in which central/satellite galaxies and dark matter haloes/subhaloes are rank ordered by their luminosities, masses, or circular velocities, and their cumulative number densities are matched.

\section*{Acknowledgments}
RAS is supported in part by the NASA \textit{Herschel} Science Center, JPL contract \#1350371, and in part by the NSF grant AST-1055081.  
RKS is supported in part by NSF 0908241 and NASA NNX11A125G, and is grateful to the GEPI and LUTH groups at Meudon Observatory for their hospitality during the summers of 2011 and 2012.  
DC acknowledges receipt of a QEII Fellowship by the Australian Research Council. 
%who shall we acknowledge?  I think it would be a good idea show the draft to Martin, Idit, Alison, and maybe Peder. 
We thank Idit Zehavi for valuable discussions about the SDSS clustering measurements. 
We thank the anonymous referee for insightful comments that helped to improve the paper.

\appendix

\renewcommand{\thefigure}{\Alph{appfig}\arabic{figure}}
\setcounter{appfig}{1}

\section{Effect of the Additional Environmental Correlation}\label{app:2alpha}

%give brief analytical description of $2\alpha$ dependence introduced to the mark CFs.
%In an appendix, we show that the expression in \ref{eq:weight} introduces an additional environmental correlation that is approximately proportional to $2\alpha$.

We provide an estimate of the effect of the added environmental correlation 
(Eqn.~\ref{eq:weight}) on the marked correlation function.

As described in Section~\ref{sec:markstats}, %{sec:MCFs}, 
the statistic $M(r)$ can be 
approximated by the simple pair count ratio $WW/DD$, where $WW$ is the sum 
over all pairs with separation $r$, weighting each member of the pair by 
its mark, and $DD$ is the total number of such pairs.  
To use a specific example in the paper (see Sec.~\ref{sec:MCFresults}), suppose 
that rather than weighting galaxies by their luminosity $L$, we modify the 
weight by adding a small dependence on the $8~\mathrm{Mpc}/h$ overdensity, 
which we will denote $1+\delta_8$.  In this case, the modified weight can 
be expressed as 
\begin{equation}
 w = L(1+\delta_8)^\alpha \sim L(1+\alpha\delta_8), 
\end{equation}
where $\alpha$ is small.  For the mark correlation function, we will 
normalize by the mean mark, and the mean of the above expression is simply 
\begin{equation}
 \langle w\rangle \sim \langle L(1+\alpha\delta_8)\rangle \sim \langle L\rangle
\end{equation}

Therefore, for a pair of galaxies $i$ and $j$ at separation $r$, we have 
\begin{equation}
 WW(r) = \frac{\langle L_i (1+\delta_i) (1+\alpha\delta_{8,i})\, L_j (1+\delta_j) (1+\alpha\delta_{8,j}) \rangle}{\langle w\rangle^2} .
\end{equation}
If $L$ is not significantly correlated with density (which is not quite 
true, because $L$ is correlated with $M_\mathrm{halo}$ and hence $\delta$), 
then this becomes 
\begin{equation}
 WW(r) = \langle [(1+\delta)(1+\alpha\delta_8)]_i \, [(1+\delta)(1+\alpha\delta_8)]_j \rangle
\end{equation}
Keeping the lowest order in $\alpha$, then this can be expanded as follows:
\begin{eqnarray}
 WW(r) &\sim& \langle (1+\delta_i)(1+\delta_j)(1+\alpha\delta_{8,i}+\alpha\delta_{8,j}) \rangle \\ %\nonumber\\
       %&\sim& \langle (1+\delta_i+\delta_j+\delta_i\delta_j)(1+\alpha\delta_{8,i}+\delta_{8,j}) \rangle \nonumber\\
       &\sim& \langle 1+\delta_i+\delta_j+\delta_i\delta_j+\alpha\delta_{8,i}(1+\delta_i+\delta_j \nonumber\\ &\phantom{\sim}&\, + \delta_i\delta_j)+\alpha\delta_{8,j}(1+\delta_i+\delta_j+\delta_i\delta_j)\rangle \nonumber\\
       %&\sim& 1 + \langle\delta_i\delta_j\rangle + \alpha(\langle\delta_{8,i}\delta_i\rangle+\langle\delta_{8,i}\delta_j\rangle+\langle\delta_{8,i}\delta_i\delta_j\rangle \nonumber\\ &\phantom{\sim}&\, + \langle\delta_{8,j}\delta_i\rangle+\langle\delta_{8,j}\delta_j\rangle+\langle\delta_{8,j}\delta_i\delta_j\rangle) \nonumber\\
       &\sim& 1 + \langle\delta_i\delta_j\rangle + 2\alpha(\langle\delta_{8,i}\delta_i\rangle+\langle\delta_{8,i}\delta_j\rangle+\langle\delta_{8,i}\delta_i\delta_j\rangle) 
\nonumber 
\end{eqnarray}

Since $DD=1+\langle\delta_i\delta_j\rangle$, we have
\begin{equation}
 \frac{WW}{DD} \sim 1 + 2\alpha \frac{\langle\delta_{8,i}\delta_i\rangle+\langle\delta_{8,i}\delta_j\rangle+\langle\delta_{8,i}\delta_i\delta_j\rangle}{DD} ,
 \label{eq:2alpha}
\end{equation}
which shows that the new environmental correlation that we introduced in the text 
produces a new signal proportional to $2\alpha$.  
The approximations here appear to be accurate, as the mark CFs with $\alpha=0.05$ can be predicted from those with $\alpha=0.01$, and vice versa, at all separations $r$ (see e.g. Fig.~\ref{fig:DCenviro}). 

%\begin{figure}
% \includegraphics[width=\hsize]{figA1.pdf} 
% \caption{caption.}
% \label{figA1}
%\end{figure}

%\section{Color-Density Relation}\label{app:colordensity}
%
%we show the color-density relation of the mock (compared to $1+\delta$-dependent colors) in Fig.~\ref{fig:colordens}. 
%% mention rescaled color dist.?
%note that the color-density for $\alpha>0$ is only slightly steepened, and would be difficult to detect. [is this useful?  if so, I will add contours to the plot.] 
%\begin{figure}
% \includegraphics[width=\hsize]{mock_Mr19_colorDCdensitymean_running10.ps} %mock_Mr19_colorDCdensitymean_widerbins.ps
% \caption{Color-density relation of the mock catalog, using DC overdensities, and $\alpha=0$ (black), 
%          0.01 (red), and 0.05 (blue).}
% \label{fig:colordens}
%\end{figure}

\section{Small-scale environment as a weight}%\label{app:envwt}

\subsection{A Toy Model}\label{app:envwt}
We will use a simple toy model to illustrate the effect of using the
environment as a weight.

Suppose that all the mass is in haloes which all have the same mass $m$
distributed around each halo centre according to
\begin{eqnarray}
 \frac{\rho(r)}{\bar\rho}
  &=& \frac{m}{\bar\rho}\,\frac{\exp(-r^2/2R_v^2)}{(2\pi R_v^2)^{3/2}} \nonumber\\
  &=& \Delta_v\,{\rm e}^{-r^2/2R_v^2}
 \label{eq:profile}
\end{eqnarray}
where the final expression defines $\Delta_v$, the central density
(because $(2\pi)^{3/2} R_v^3$ is the volume of the profile).

Then the unweighted correlation function is
\begin{eqnarray}
 \xi(r) &=& \frac{\bar\rho}{m}\, \frac{m^2}{\bar\rho^2}\,
                \frac{\exp(-r^2/4R_v^2)}{(2\pi\, 2R_v^2)^{3/2}} \nonumber\\
  &=& \frac{\Delta_v}{2^{3/2}}\,{\rm e}^{-r^2/4R_v^2},
\end{eqnarray}
where the first factor of $\bar\rho/m$ is the number density of haloes
(i.e., all the mass is in haloes of mass $m$).
%On small scales, this tends to $\Delta_v/2^{3/2}$.

If we model the weight as the local value of the density smoothed with a
fixed aperture of scale $s$, then the weight associated with a distance
$r$ from the halo centre is
\begin{equation}
 w(r) = {\rm e}^{-r^2/2(R_v^2 + s^2)}.
\end{equation}
If we define
\begin{equation}
 R_s^2 \equiv R_v^2 \, (R_v^2 + s^2)/(2R_v^2+s^2)
\end{equation}
then the mean weight is
\begin{equation}
 \bar w = 4\pi \int dr\,r^2\,\rho(r)\,w(r)/m
 %       = 4\pi \int dr\,r^2\,\frac{{\rm e}^{-r^2/2R_s^2}}{(2\pi R_v^2)^{3/2}}
 %       = (R_v^2 + s^2)^{3/2}/(2R_v^2  + s^2)^{3/2}
        = (R_s/R_v)^3.
\end{equation}
Therefore, the normalized weighted correlation function is
\begin{equation}
 \xi_{w}(r) = \frac{\bar\rho}{m}\, \frac{m^2}{\bar\rho^2}\,
               \frac{{\rm e}^{-r^2/4R_s^2}}{(2\pi\, 2R_s^2)^{3/2}}
\end{equation}
To see what this implies for WW/DD, suppose that $s\ll R_v$.  Then
$R_s^2\to R_v^2/2$.  On small scales
 WW/DD $= (1 + \xi_w)/(1 + \xi) \approx \xi_w/\xi$ because we are
interested in the case in which $\Delta_v\gg 1$.  In this limit,
 WW/DD $\approx 2^{3/2}\,\exp(-r^2/2R_v^2)/\exp(-r^2/4R_v^2)$.
This shows that WW/DD has the same shape as $\xi$ itself,
and the small-scale amplitude is $2^{3/2}$ times the unweighted one.
The amplitude is $2^{3/2}$ because the assumed Gaussian profile (Eqn.~\ref{eq:profile}) is 
relatively flat; a centrally cusped profile [such as a Navarro, Frenk, \& White (1996) one] will produce a stronger signal. 
The small-scale shape of WW/DD 
should not come as a surprise:  when $s\ll R_v$ then $\xi_w$
is like the convolution of $\rho^2$ with itself, making
$\xi_w\propto \xi^2$.
More generally, WW/DD $\propto\xi^{1/(1 + s^2/R_v^2)}$.

\subsection{A Model in terms of Cluster and Field Populations}\label{app:clusterfield}

To gain intuition about the effect of rank ordering, first note that 
for a list of length $N$ marks, the mean mark is
 $[N(N+1)/2]/N = (N+1)/2$, 
so normalizing is particularly simple.  Now, suppose the distribution 
of environments were bimodal, with one population associated with 
`close' pairs (separations less than some $R_c$) in dense regions, 
and another with underdense ones.  
%:  our procedure would decrease the signal.  
%in the following, `reds' & `blues' were changed to `cluster' & `field'
Suppose that `cluster' galaxies have close neighbours but `field' galaxies do 
not (i.e. they are like hard spheres).  
Then the total clustering signal is 
 $n_t^2(1+\xi_{tt}) = n_c^2(1+\xi_{cc}) + n_f^2(1+\xi_{ff}) + 2n_cn_f(1+\xi_{cf})$,
where $n_t\equiv n_c + n_f$ and the mean mark is 
$\bar w \equiv (n_cw_c + n_fw_f)/n_t$.  
Therefore, 
\begin{equation}
\frac{WW}{DD} = (n_cw_c)^2(1+\xi_{cc}) + (n_fw_f)^2(1+\xi_{ff}) + 
 \frac{2n_cw_cn_fw_f(1+\xi_{cf})}{\bar w^2 n_t^2(1+\xi_{tt})} .
\end{equation}
If we have rank ordered the marks, then 
$n_t \bar w = N(N+1)/2$, $n_f w_f = N_f(N_f+1)/2$, and 
$n_c w_c = n_t \bar w - n_f w_f$.  
On scales smaller than $R_c$ we know that both $\xi_{ff}$ and 
$\xi_{cf}$ equal $-1$, making
\begin{eqnarray}
 \frac{WW}{DD} &=& \frac{n_c^2 w_c^2(1+\xi_{cc})}{n_c^2 \bar w^2(1+\xi_{cc})} = \biggl(\frac{w_c}{\bar w}\biggr)^2 \nonumber\\
        &=& [1 + N_f/(N+1)]^2 . 
\end{eqnarray} 
Thus, the small scale signal is a measure of the field fraction $N_f/N$, but notice that it cannot exceed 4.  
%:  our procedure would decrease the signal.  

If we were to interpret our measured value of $WW/DD=3$ in Figure~\ref{fig:densityMCFs} in these terms, we would infer a field fraction of about 70\%; 
it is interesting that this implies a cluster fraction (30\%) that is close to the satellite fraction usually quoted in halo model analyses of galaxy clustering (e.g. Zehavi et al.\ 2005; van den Bosch et al.\ 2007) 
and the satellite fraction of the mock catalogue used in this paper (Sec.~\ref{sec:mock}).  
If we assume that on intermediate scales $\xi_{ff}$ and $\xi_{cf}$ are both approximately equal to zero, then 
\begin{equation}
 \frac{WW}{DD} = \frac{1 + (n_c/n_t)^2 (w_c/\bar w)^2\xi_{cc}}{1 + (n_c/n_t)^2\xi_{cc}} . 
\end{equation}
In this approximation, the scale dependence of $WW/DD$ codes information about the cluster or field fraction, and the correlation function of the cluster population.  In the $\xi(r)\gg 1$ limit, it smoothly asymptotes to the previous expression.  

\label{lastpage}

\end{document}